\documentclass{aa}  
\bibpunct{(}{)}{;}{a}{}{,}

\usepackage{graphicx}

\usepackage{txfonts}

\usepackage{longtable}
\usepackage{amsmath}

\usepackage{pdflscape}
\usepackage{booktabs}
\usepackage{adjustbox}
\usepackage{threeparttable}

\usepackage{soul}

\usepackage{fixmath}

\usepackage[colorlinks=true,linkcolor=blue,citecolor=blue]{hyperref}

\newcommand{\Autoref}[1]{%
  \begingroup%
  \def\chapterautorefname{Chapter}%
  \def\sectionautorefname{Sect.}%
  \def\subsectionautorefname{Sect.}%
  \def\equationautorefname{Eq.}%
  \def\figureautorefname{Fig.}%
  \autoref{#1}%
  \endgroup%
}

\usepackage{xcolor}

\newcommand{\Eqref}[1]{%
  \hyperref[#1]{\textcolor{blue}{Eq.~(\ref*{#1})}}%
}

\def \angstrom{$\text{\AA}$ }

\newcommand{\EqParRef}[1]{%
  \hyperref[#1]{\textcolor{blue}{(\ref*{#1})}}%
}

\begin{document} 

 \title{The GAPS Programme at TNG\thanks{Based on observations made with the Italian \textit{Telescopio Nazionale Galileo} (TNG), operated on the island of La Palma by the INAF - \textit{Fundaci\'on Galileo Galilei} at the \textit{Roque de Los Muchachos} Observatory of the \textit{Instituto de Astrof\'isica de Canarias} (IAC).}} 
 
 \subtitle{LXXX. Characterisation of the 400 Myr old multi-planet system K2-233}
   
   \titlerunning{The GAPS Programme at TNG: the young multi-planet system K2-233}
   \authorrunning{V. Fardella et al.}

   \author{V. Fardella\inst{1,2} \corrauth{valerio.fardella@inaf.it},
           S. Benatti\inst{2} \email{serena.benatti@inaf.it},
           R. Spinelli\inst{2} \email{riccardo.spinelli@inaf.it},
           D. Locci\inst{2} \email{daniele.locci@inaf.it},
           G. King\inst{3,4,5} \email{kinggw@umich.edu},
           M. Baratella\inst{6} \email{Martina.Baratella@eso.org},
           K. Biazzo\inst{7} \email{katia.biazzo@inaf.it},
           S. Desidera\inst{8} \email{silvano.desidera@inaf.it},
           D. Barbato\inst{8} \email{domenico.barbato@inaf.it},
           M. C. D'Arpa\inst{2} \email{mattia.darpa@inaf.it},
           L. Naponiello\inst{9} \email{luca.naponiello@inaf.it},
           A. Petralia\inst{2} \email{antonino.petralia@inaf.it},
           M. Damasso\inst{9} \email{mario.damasso@inaf.it},
           O. Barragán\inst{17, 18} \email{oscar.barragan@warwick.ac.uk},
           C. Di Maio\inst{2} \email{claudia.dimaio@inaf.it},
           A. Maggio\inst{2} \email{antonio.maggio@inaf.it},
           G. Piccinini\inst{2, 1} \email{giulia.piccinini@inaf.it}, 
           A. Bonomo\inst{9} \email{aldo.bonomo@inaf.it}, 
           I. Carleo\inst{9} \email{ilaria.carleo@inaf.it},
           R. Claudi\inst{8,10} \email{riccardo.claudi@inaf.it},
           M. Gonzales\inst{11} \email{manuel.gonzalez@inaf.it},
           A. Harutyunyan\inst{11} \email{avet@tng.iac.es},
           A. F. Lanza\inst{12} \email{antonino.lanza@inaf.it},
           L. Malavolta\inst{13, 8} \email{luca.malavolta@unipd.it},
           L. Mancini\inst{14} \email{lmancini@roma2.infn.it},
           G. Mantovan\inst{14, 8} \email{giacomo.mantovan@unipd.it},
           D. Nardiello\inst{13, 8} \email{domenico.nardiello@unipd.it},
           M. Rainer\inst{16} \email{monica.rainer@inaf.it},
           A. Sozzetti\inst{9} \email{alessandro.sozzetti@inaf.it},
           T. Zingales\inst{13} \email{tiziano.zingales@unipd.it}
           }

   \institute{
        University of Palermo, Department of Physics and Chemistry “Emilio Segrè”, Via Archirafi 36, Palermo, Italy.
        \and
        INAF – Osservatorio Astronomico di Palermo, Piazza del Parlamento, 1, 90134 Palermo, Italy.
        \and
        Department of Astronomy, University of Michigan, Ann Arbor, MI 48109, USA.
        \and
        Department of Physics, University of Warwick, Gibbet Hill Road, Coventry CV4 7AL, UK.
        \and
        Centre for Exoplanets and Habitability, University of Warwick, Gibbet Hill Road, Coventry CV4 7AL, UK.
        \and
        ESO - European Southern Observatory, Alonso de Córdova 3107, Casilla 19, Santiago 19001, Chile.
        \and
        INAF – Osservatorio Astronomico di Roma, Via Frascati 33, 00040 Monte Porzio Catone (RM), Italy.
        \and
        INAF - Osservatorio Astronomico di Padova, Vicolo dell’Osservatorio 5, I-35122 Padova, Italy.
        \and
        INAF – Osservatorio Astrofisico di Torino, Via Osservatorio 20, 10025 Pino Torinese, Italy.
        \and 
        Dipartimento di Matematica e Fisica, Università Roma Tre, Via della Vasca Navale 84, 00146 Roma, Italy
        \and
        Fundación Galileo Galilei - INAF, Rambla José Ana Fernandez Pérez 7, E-38712 Breña Baja (TF), Spain.
        \and
        INAF – Osservatorio Astrofisico di Catania, Via S. Sofia 78, 95123 Catania, Italy.
        \and
        Dipartimento di Fisica e Astronomia "Galileo Galilei", Università di Padova, Vicolo dell’Osservatorio 3, I-35122 Padova, Italy.
        \and
        Department of Physics, University of Rome “Tor Vergata”, Via della Ricerca Scientifica 1, 00133 Rome, Italy.
        \and
        Centro di Ateneo di Studi e Attività Spaziali "G. Colombo" -- Università degli Studi di Padova, Via Venezia 15, IT-35131, Padova, Italy.
        \and
        INAF – Osservatorio Astrofisico di Arcetri, Largo Enrico Fermi 5, 50125 Firenze, Italy.
        \and
        Department of Physics, University of Warwick, Coventry, CV4 7AL, UK.
        \and
        Sub-department of Astrophysics, Department of Physics, University of Oxford, Oxford, OX1 3RH, UK.
        }

   \date{}

  \abstract
   {The study of the physical properties of the exoplanets is crucial to retrieve the characteristics of their atmospheres and their inner composition. Precise measurements of exoplanetary masses are crucial to constrain their bulk densities and infer their inner composition. For young systems hosting planets within or near the radius valley, accurate physical parameters are required to distinguish between different evolutionary scenarios and to understand the impact of atmospheric mass-loss mechanisms, such as photoevaporation driven by stellar activity.}
   {In this work we perform a revision of the masses of the three transiting exoplanets hosted by K2-233, a young ($\sim 400$ Myr old) K-type active star. Through dedicated simulations, we get insights the planets' atmospheres evolution with time affected by photoevaporation, and into the modelling of the planets' interior.}
   {We collected 70 HARPS-N observations within the GAPS/ArMS Large Programme and performed a joint analysis combining 124 HARPS literature observations, the available K2 light curves, and TESS data obtained simultaneously with our radial velocity monitoring. After obtaining the planetary information, we reconstructed the atmospheric evolution of the planet K2-233 d by using our photoevaporation code.}
   {Due to the strong stellar activity of K2-233, we explored several Multi-GP configurations and demonstrated that the inferred planetary masses are highly sensitive to the selected activity-modelling strategy, yielding a set of different physically viable solutions rather than a single uniquely preferred model.
   Our models indicate that K2-233 d likely experienced significant atmospheric mass loss. In most scenarios, the planet evolves toward a stripped or partially stripped state, placing it within or near the radius valley at Gyr timescales. Depending on its internal composition and atmospheric properties, the final outcome ranges from a low-mass sub-Neptune retaining a thin envelope ($R_p \sim 1.71 - 1.84 \ R_{\oplus}$) to a bare-core super-Earth ($R_p \sim 1.95 \ R_{\oplus}$). From our inner composition models, we found that K2-233 b and K2-233 c are likely iron-dominated, while K2-233 d presents a significant H/He atmosphere.}
   {}


   \maketitle
   \nolinenumbers

\section{Introduction}

The study of young systems is crucial for understanding planetary formation and evolution. The Kepler mission \citep{Borucki_2010}, its extension K2 \citep{Howell_2014}, and the Transiting Exoplanet Survey Satellite (TESS; \citealt{tess_2015}) have discovered hundreds of transiting exoplanets using the transit photometry (TP) technique, including several planets around young stars (e.g. \citealt{Mann_2016}; \citealt{Rizzuto_2020}; \citealt{Tran_2021}; \citealt{Chen_2021}; \citealt{Zakhozhay_2022}; \citealt{Filomeno_2024}; \citealt{nardiello2025}; \citealt{Filomeno_2026}). These observations provide measurements of planetary radii and orbital periods. Planetary masses can instead be obtained through the radial velocity (RV) method using high-resolution spectrographs such as the High Accuracy Radial velocity Planet Searcher (HARPS; \citealt{Mayor_2003}) at the European Southern Observatory (ESO) and the High Accuracy Radial velocity Planet Searcher for the Northern hemisphere (HARPS-N; \citealt{Cosentino_2012_harpsn}) at the Telescopio Nazionale Galileo (TNG).
When both TP and RV measurements are available, the mean density of a planet can be derived. However, density alone cannot uniquely determine the planetary composition because different internal structures can produce similar bulk properties \citep{Zeng_2019}. This degeneracy is particularly relevant for planets located in the so-called radius valley (\citealt{Owen_2013}; \citealt{Fulton_2017}) at approximately $R_p \sim 1.8$ $R_\oplus$, which separates super-Earths, whose radius distribution peaks at $\sim 1.3$ $R_\oplus$ from sub-Neptunes, peaking at $\sim 2.4$ $R_\oplus$. One proposed explanation for this feature is atmospheric mass loss driven by intense X-ray and extreme-ultraviolet (XUV) irradiation from the host star, which can strip hydrogen- and helium-rich envelopes through photoevaporation (\citealt{Lopez_2013}; \citealt{Owen_2013}). In this scenario, sub-Neptunes retain part of their gaseous envelopes, resulting in larger radii.
An alternative explanation is core-powered mass loss, where the cooling luminosity of the planetary core drives hydrodynamic escape and gradually removes light elements from the primordial atmosphere \citep{Ginzburg_2018}. A third scenario proposed by \cite{Luque_2022} suggests that planets in the radius valley orbiting M-type stars may be water-rich worlds that formed beyond the snow line and later migrated inward. 
Although this mechanism is expected to be more relevant for low-mass stars, it highlights the broader diversity of possible evolutionary pathways for planets in the radius valley.
Distinguishing between these mechanisms requires atmospheric characterisation to break the degeneracy in planetary composition. 
Current facilities such as the James Webb Space Telescope (JWST; \citealt{Gardner_2006_jwst}) are already enabling detailed atmospheric studies of exoplanets, while in the future, the Ariel Space Mission \citep{Tinetti_2018} will observe the atmospheres of more than $1000$ exoplanets, from rocky planets to gas giants, to investigate their formation and evolution (\citealt{Zingales2018ArielMRS}; \citealt{Edwards2019ArielTargets}).
Atmospheric studies rely on accurate knowledge of planetary parameters, particularly the mass \citep{Di_Maio_2023}. The Ariel Masses Survey (ArMS) Large Programme (\citealt{Bonomo_2025}; \citealt{Filomeno_2026}; \citealt{DiMaio_2026}), conducted within the Global Architecture of Planetary Systems (GAPS) collaboration (\citealt{Covino_2013}; \citealt{Carleo_2020}), uses the HARPS-N spectrograph at TNG to obtain high-precision spectra for systems hosting planets within or near the radius valley, enabling robust mass measurements through the RV technique.

For young and active stars, both TP and RV measurements can be affected by stellar activity, which introduces spurious signals through photometric variability and distortions of spectral lines (\citealt{Damasso_2020}; \citealt{DiMaio_2024_spotccf}). Gaussian Processes (GPs; \citealt{Aigrain_2012}) are widely used to model such activity-induced variations. Their performance depends on the adopted covariance function, with the quasi-periodic (QP) kernel (\citealt{Nicholson_2022} and references therein) being commonly used because it captures the periodic behaviour associated with stellar rotation. More recently, multi-dimensional GP models (Multi-GPs; \citealt{Rajpaul_2015}) have been applied to RV analyses by jointly modelling RVs and spectroscopic activity indicators, often improving the recovery of planetary signals \citep{Rajpaul_2015}.

In this work we investigate the K2-233 system, which hosts three transiting planets with orbital periods of approximately $2.4$, $7$, and $24$ days \citep{David_2018} orbiting a young ($\sim 400$ Myr) and active K-type star. The system has been studied previously. Using a one-dimensional GP, \cite{2020lillo-box} derived upper limits on the planetary masses using a one-dimensional GP, obtaining $11.26$ $M_\oplus$, $12.81$ $M_\oplus$, and $21.14$ $M_\oplus$ for planets b, c, and d. \cite{Barrag_n_2023} applied a Multi-GP approach, recovering partial mass detections for planets b and c and a robust measurement for planet d: $2.4^{+1.5}_{-1.3}$ $M_\oplus$, $4.6^{+1.8}_{-1.7}$ $M_\oplus$, and $10.3^{+2.4}_{-2.6}$ $M_\oplus$. \cite{Osborne_2024} re-analysed a sample of HARPS targets, including K2-233, finding masses of $2.13^{+1.55}_{-1.33}$ $M_\oplus$, $6.96^{+1.96}_{-1.99}$ $M_\oplus$, and $4.71^{+2.74}_{-2.58}$ $M_\oplus$ for planets b, c, and d, respectively.
The discrepancies among these measurements highlight the sensitivity of RV-derived planetary masses to the modelling of stellar activity. K2-233 has been included in the ArMS target list, since it was initially selected for the Ariel Mission Candidate Sample \citep{Edwards_2022}.
The masses of the planets of K2-233 retrieved from the previous studies are not in full agreement between them. Since robust mass determinations are essential to properly model internal compositions and evolutionary pathways – especially for young systems where intense stellar activity heavily contaminates the signals – here we present a new analysis of K2-233 combining the new RV measurements with archival data and recent TESS photometry with partial overlap with the HARPS-N monitoring. This overlap allows us to perform a joint GP analysis of RV and light-curve data in order to better characterise the stellar variability.
The paper is organised as follows. \Autoref{sec2:observations} describes the observations, \Autoref{sec3:stellar_analysis} presents the stellar characterisation, \Autoref{sec:data-analysis} details the analysis of the planetary signals, \Autoref{sec:photoevaporation} describes the modelling of atmospheric photoevaporation, and \Autoref{sec:discussion} shows the modelling of the planets' interior.
The discussions and conclusions are presented in \Autoref{sec:conclusions}.

\section{Observations}
\label{sec2:observations}

In this section, we describe the photometric and spectroscopic datasets employed in this work, including the preparation and quality control of the data, which form the basis of our joint analysis. 

\subsection{Photometric observations}

\begin{figure*}
\centering
    \includegraphics[width=0.95\textwidth]{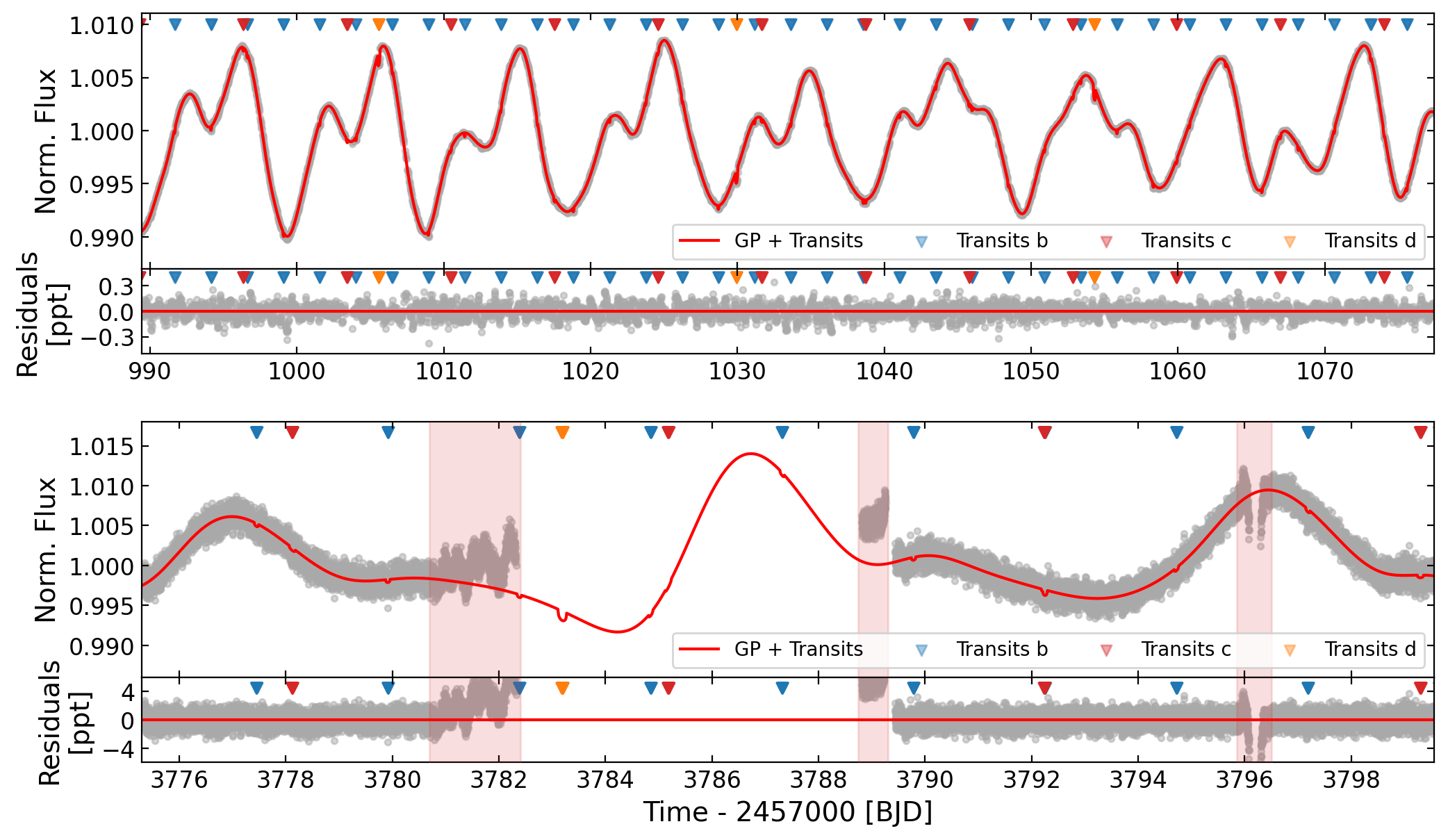}
     \caption{Normalised K2 photometric time series of K2-233 (top panel) and TESS observations (bottom panel). The red lines show the models obtained from the Gaussian Processes combined with the transit signals of the three planets. As indicated in the legend, coloured triangles mark the transit times of each planet. Note that the only transit of planet d expected in the TESS dataset (at $\sim 3783$ days $- 2457000$ BJD) falls within a data gap; therefore, no transit of planet d is actually observed. The pink shaded regions highlight portions of the time series affected by residual systematics, which were excluded from the analysis.}
    \label{sec:data:KeplerTESSLC}
\end{figure*}

K2-233 was first observed by the K2 mission between August 23rd and November 20th 2017 with a cadence of 30 minutes. Using these data, \cite{David_2018} detected and analysed the system’s three transiting planets, deriving the orbital periods $P_b = 2.46746 \pm 0.00014$ days, $P_c = 7.06142 \pm 0.00084$ days, $P_d = 24.3662 \pm 0.0021$ days and radii $R_b = 1.398^{+0.062}_{-0.060}$ $R_\oplus$, $R_c = 1.335^{+0.083}_{-0.077}$ $R_\oplus$ and $R_d = 2.64 \pm 0.11$ $R_\oplus$ for planets b, c, and d, respectively.
We downloaded the K2 light curve of K2-233 using the \texttt{EVEREST} Python package (\citealt{Luger_2016}; \citealt{Luger_2018}), which also provided a correction for instrumental systematics.
After this correction, we masked the transits of the planets and performed a sigma clipping at $3.5\sigma$ to remove the outliers. The transits were then reinserted for subsequent analysis.
The top panel of \Autoref{sec:data:KeplerTESSLC} displays the corrected K2 light curve, showing a clear modulation induced by the stellar activity, which has been modelled using a GP (see \Autoref{sec:data-analysis}). 

K2-233 was also observed by TESS from April 9th to May 7th 2025 using a cadence of 120 seconds. We downloaded the data using the \texttt{lightkurve} Python package \citep{lightkurve_2018}. In the bottom panel of \Autoref{sec:data:KeplerTESSLC} we show the Pre-search Data Conditioning Simple Aperture Photometry (PDC-SAP) light curve from TESS Sector 91, which is also what we used for the analyses in this work. The PDC-SAP presents three regions contaminated by residual instrumental systematics at approximately $2460780.7$ BJD (2025 April 15th 4:48 UT), $2460788.5$ BJD (2025 April 23rd 0:00 UT), and $2460795.85$ BJD (2025 April 30th 8:24 UT), highlighted in \Autoref{sec:data:KeplerTESSLC}. These segments were removed to prevent spurious effects in the analysis.
As the light curve collected by K2, also the light curve of TESS presents a modulation produced by the stellar activity that has been modelled through a GP in \Autoref{sec:data-analysis}.
We also note that TESS did not observe any transit of planet d during this sector. As for the K2 dataset, we applied a $3.5\sigma$ sigma clipping to remove outliers.

\subsection{Spectroscopic observations}
\label{sec:obsv:spectr}

\begin{figure*}
\centering
    \includegraphics[width=\textwidth]{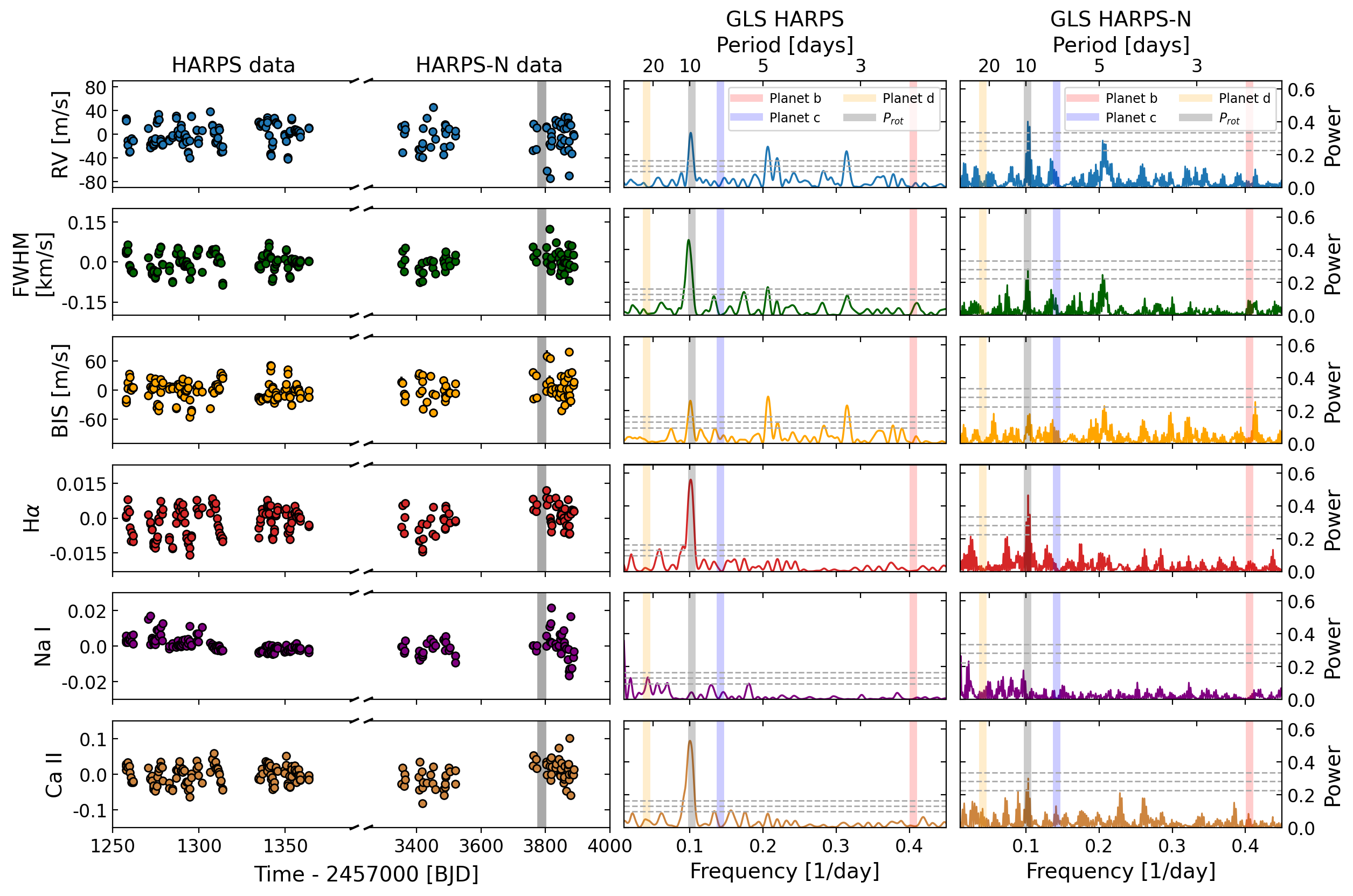}
     \caption{Time series (on the left column) and Generalised Lomb-Scargle (GLS) periodograms of the RV and the activity indicators of the K2-233's spectra collected by HARPS (central column) and HARPS-N (right column). Starting from the top the panels show the RVs, and the BIS of the CCF, and the H$\alpha$, \ion{Na}{i}, and \ion{Ca}{ii} of the spectra. The error bars in the time series are smaller than the dots size that represent the values of the data.
     The grey vertical areas indicate the time window within the TESS sector 91 was collected.
     In the GLS periodograms, starting from the bottom, the three horizontal dashed lines represent the False Alarm Probability (FAP) at $10\%$, $1\%$, and $0.1\%$.}
    \label{sec:data:GLS}
\end{figure*}

K2-233 was observed in the framework of the GAPS/ArMS programme using the HARPS-N spectrograph mounted at the $3.6$ m Telescopio Nazionale Galileo (TNG) as part of the GAPS collaboration. We collected 70 spectra spanned in two semesters: from 2024 February 12th to 2024 July 28th (167 days), and from 2025 March 26th to 2025 July 30th (126 days). We note that the second semester of HARPS-N observations overlapped with the TESS photometric campaign. 
We downloaded the data from the YABI workflow hosted at the IA2 Data Center\footnote{\url{https://www.ia2.inaf.it/}} of the INAF-Astronomical Observatory of Trieste. We processed the HARPS-N data using the official HARPS-N DRS v$3.7$ by cross-correlating the spectra with a K5 numerical mask to obtain the cross-correlation function (CCF, see \citealt{Pepe_2002} and references therein) from which we extracted the radial velocities as well as the corresponding full width half maximum (FWHM) and the bisector inverse slope (BIS). 
Since the presence of planets does not affect the stellar FWHM or BIS, their temporal variations reflect intrinsic stellar activity. 
The procedures available on YABI also allow to extract the $\log R'_{\text{HK}}$ of the spectra using the method described in \cite{Lovis_2011}, from which we estimated an average value of $ \langle\: \log R'_{\text{HK}} \:\rangle = -4.38^{+0.03}_{-0.02}$.
We also extracted the H$\alpha$, \ion{Na}{i}, and \ion{Ca}{ii} activity indicators with the \texttt{ACTIN 2} Python package (\citealt{Gomes_2018}; \citealt{Gomes_2021}).

The HARPS-N spectra have a median Signal-to-Noise Ratio (SNR) of 42 at 550 nm, with a median RV uncertainty of $2.7$ m s$^{-1}$. We removed the spectra with $\text{SNR} < 25$, for which the RV uncertainty exceeded 5 $\text{m s}^{-1}$, resulting in the exclusion of six observations\footnote{2024 February 16th 2024 06:37 UT, 2024 July 12th 23:54 UT, 2025 April 4th 02:39 UT, 2025 April 4th 03:01 UT, 2025 July 22nd 20:44 UT and 2025 July 23rd 20:52 UT.} .
The dispersion of the remaining 64 RVs collected from HARPS-N is $24$ $\text{m s}^{-1}$.

K2-233 was also observed with the HARPS spectrograph (Programme ID 198.C-0169, PI: Santerne) mounted on the 3.6 m ESO telescope at La Silla Observatory, Chile. A total of 124 high-resolution spectra were acquired over a baseline of 105 days. 
We downloaded the HARPS spectra from the ESO archive, and we reduced them using the dedicated HARPS DRS v$3.8$, again using a K5 numerical mask to obtain the CCF and subsequently the RV, BIS, and FWHM values. 
We additionally extracted the H$\alpha$, \ion{Na}{i}, and \ion{Ca}{ii} chromospheric indicators using \texttt{ACTIN 2}.

These spectra have a median SNR of 51 (at 550 nm), with a median RV uncertainty of $1.82$ m s$^{-1}$.
Two spectra exhibit $\text{SNR} < 25$, resulting in RV uncertainties exceeding $5$ m s$^{-1}$\footnote{2018 August 5th 2:28 UT and 2018 August 8th 0:42 UT.}. We decided to discard them from the analyses in order to ensure more accurate results.
The dispersion of the RV collected from HARPS is $18$ $\text{m s}^{-1}$, slightly smaller than that of the spectra collected by HARPS-N.

\Autoref{sec:data:GLS} shows the RV and activity-indicator time series for both HARPS and HARPS-N, together with their Generalised Lomb–Scargle (GLS) periodograms \citep{Zechmeister_2009}. With the exception of the \ion{Na}{i} index, all GLS periodograms exhibit a dominant peak close to the stellar rotation period, $P_{\text{rot}} = 9.754$ days, as reported by \cite{David_2018}.
The RV and BIS periodograms also display peaks at $\sim 4.9$ and $\sim 3.3$ days, corresponding to the first and second harmonics of the stellar rotation signal.
The \ion{Na}{i} lines form in the upper photosphere and lowermost chromosphere, whereas H$\alpha$ and \ion{Ca}{ii} lines originate in the upper chromosphere. As a result, the \ion{Na}{i} index is generally considered a reliable activity indicator primarily for cool M-type stars \citep{Gomes_2011_NaI}.
No significant power is detected at the planetary periods in the RV GLS, confirming that the RV time series is dominated by stellar activity.

\subsection{XMM-Newton observations}
\label{subsec:xmm-newton-data}

We observed K2-233 with XMM-Newton on August 1st/2nd 2023 (PI: King; ObsID: 0922560201). We reduced the data from the three European Photon Imaging Cameras (EPIC) using SAS 22.1.0, using standard procedures\footnote{As outlined on the SAS Threads: \url{https://www.cosmos.esa.int/web/xmm-newton/sas-threads}}. This included filtering for periods of high solar proton background \citep{Walsh2014}, which especially affected much of the second half of the observation, leaving us with $7.6$, $23.1$, and $24.0$\,ks of good time intervals for EPIC-pn, EPIC-MOS1, and EPIC-MOS2, respectively. The source was still clearly detected by all three cameras. For source extraction, we used 15 arcsec regions centred on the proper motion corrected position of the star.

\section{Stellar analysis}
\label{sec3:stellar_analysis}

In this section we characterise the host star K2-233 using the available spectroscopic and photometric information. 
First, we derive the stellar atmospheric parameters and chemical abundances from the HARPS-N coadded spectrum. 
We then combine these results with additional activity and kinematic indicators to estimate the stellar age, mass, and radius.

\subsection{Spectroscopic stellar parameters}
\label{sec:stellar}

    \begin{table}
        \centering
        \caption[]{Main parameters of K2-233.}
        \label{tab:K2-233_parameters}
        \begin{tabular*}{\columnwidth}{@{\extracolsep{\fill}}lcc}
            \toprule
            Parameter & Value & Source \\
            \midrule
            \multicolumn{3}{l}{$\!\!\!$\textit{Astrometric properties}} \\
            R.A., J2000 [hh:mm:ss] & $15$:$21$:$55.198$ & [1] \\[4pt]
            Decl., J2000 [dd:mm:ss] & $-20$:$13$:$53.991$ & [1] \\[4pt]
            Parallax [mas] & $14.7719 \pm 0.0188$ & [1] \\[4pt]
            Distance [pc] & $67.696 \pm 0.0862$ & [1] \\
            
            \multicolumn{3}{l}{} \\
            \multicolumn{3}{l}{$\!\!\!$\textit{Photometric properties}} \\
            $B$ [mag] & $11.664 \pm 0.027 $ & [2] \\[4pt]
            $V$ [mag] & $10.726 \pm 0.019$ & [2] \\[4pt]
            $J$ [mag] & $8.968 \pm 0.020$ & [3] \\[4pt]
            $H$ [mag] & $8.501 \pm 0.026$ & [3] \\[4pt]
            $K_s$ [mag] & $8.375 \pm 0.023 $ & [3] \\
            
            \multicolumn{3}{l}{} \\
            \multicolumn{3}{l}{$\!\!\!$\textit{Spectroscopic properties}} \\

            $T_{\text{eff}}$ [K] & $4885 \pm 50$ & [4] \\[4pt]
            $\log g$ $ [\text{cm s$^{-2}$}]$ & $4.58 \pm 0.05$ & [4] \\[4pt]
            $V_t$ [km s$^{-1}$] & $0.71\pm0.25$ & [4] \\[4pt]
            $[\text{Fe/H}]$ [dex] & $-0.03 \pm 0.05$ & [4] \\[4pt]
            $[\text{Mg/H}]$ [dex] & $-0.10 \pm 0.08$ & [4] \\[4pt]
            $[\text{Si/H}]$ [dex] & $-0.01 \pm 0.08$ & [4] \\[4pt]
            Age [Myr] & $400^{+200}_{-150}$ & [4] \\[4pt]
            $P_{\text{rot}}$ [days] & $9.613^{+0.055}_{-0.059}$ & [4] \\[4pt]
            $v \sin i$ [km s$^{-1}$] & $4.5 \pm 1.0$ & [5] \\[4pt]
            $ \langle\: \log R'_{\text{HK}} \:\rangle$ & $-4.38^{+0.03}_{-0.02}$ & [4] \\[4pt]
            $L$ [$L_\odot$] & $0.244 \pm 0.010$ & [4] \\[4pt]
            Mass [$M_\odot$] & $0.795 \pm 0.009$ & [4] \\[4pt]
            Radius [$R_\odot$] & $0.692 \pm 0.004$ & [4] \\[4pt]
            Density [$\text{g cm}^{-3}$] & $3.38 \pm 0.07$ & [4] \\
            \noalign{\smallskip}
            \bottomrule
        \end{tabular*}
        \tablebib{
            [1] \textit{Gaia} DR3; [2] APASS DR9; [3] 2MASS; [4] This work; [5] \cite{David_2018}.
        }
   \end{table}

Following the same procedure as extensively described in \cite{2022Biazzo}, we analysed the HARPS-N coadded spectrum (SNR $\sim 450$ at 550 nm) of our target adopting the standard equivalent width (EW) method. We used in combination the {\sc ARES} (v2, \citealt{ares}) and {\sc pymoogi} \citep{pymoogi} codes, together with the ATLAS9+ODFNEW model atmospheres \cite{castellikurucz2003}, to measure the EW of \ion{Fe}{I} and \ion{Fe}{II} lines and derive temperature ($T_{\mathrm{eff}}$), gravity ($\log g$) and micro-turbulence velocity parameter ($V_t$). We excluded from the analysis lines with uncertainties in the EW measurements larger than 10$\%$ and we checked manually with IRAF those with EW > 120 mA. The final spectroscopic stellar parameters are derived by imposing three different equilibria simultaneously: the excitation and ionisation equilibria (to derive $T_{\mathrm{eff}}$ and $\log g$ respectively), and zeroing the trend between reduced EW of \ion{Fe}{I} lines and their abundances. The results are reported in \Autoref{tab:K2-233_parameters}.

The spectroscopic $T_{\mathrm{eff}}$ is in excellent agreement with the estimates obtained with the tool {\sc colte} \citep{colte} using 12 different colour indexes with 2MASS \citep{2MASS} and \textit{Gaia} Data Release 3 (DR3) photometry \citep{gaia}. The photometric temperatures span from a minimum value of $4817\pm50$\,K in $(G-B_P)$ to a maximum of $4908\pm67$\,K in $(R_P-H)$, with a mean value of $4860\pm48$\,K. Our inferred spectroscopic gravity confirms what we derived using the classical equation with the absolute luminosity and the mass equal to $0.78\pm0.09$ $M_{\odot}$ taken from TIC \citep{tic}. This input gravity value is $4.61\pm0.09$ dex, well in agreement also with the \textit{Gaia} measurement of $4.56$ dex. Finally, the derived micro-turbulence velocity agrees with the value derived with the relation published in \cite{2016dutra} of $0.71\pm0.03$\,km\,s$^{-1}$. Our analysis confirms also the findings by \cite{2020lillo-box}, where the authors applied a similar methodology but with a different line list, which can explain the difference in temperature and metallicity. 

We also computed the elemental abundances of magnesium ([Mg/H]) and silicon ([Si/H]) using an EW-based approach, and adopting the same codes, model atmospheres, line list, and analysis procedure described in \cite{2022Biazzo}. The abundances were derived from a set of carefully selected and unblended \ion{Mg}{i} and \ion{Si}{i} lines, and were computed under the assumption of local thermodynamic equilibrium. For both elements, the final abundance was obtained differentially with respect to the solar values. For further details, we refer the reader to the reference mentioned above.

\subsection{Stellar age, mass, and radius}
\label{subsec:age-mass-radius}

For the determination of the age of the system, we considered a variety of indicators.
As previously discussed in the literature \citep{David_2018}, and shown in the analysis presented in \Autoref{sec2:observations}, K2-233 has prominent photometric and spectroscopic variability supporting a moderately young age.
For the adopted rotation period (see \Autoref{subsec:lc_analysis}), the resulting age is slightly older than Group X \citep[about 300 Myr, ][]{messina2022} and clearly younger than the Hyades and Praesepe open clusters.
The chromospheric emission ($\langle \,\log R'_{\text{HK}} \,\rangle = -4.38^{+0.03}_{-0.02}$, the average value extracted from the HARPS-N spectra) and the X-ray luminosity from \Autoref{sec:x-ray} are both slightly above the mean value of Hyades stars of similar colour, supporting a younger age. The value of $\log R'_{\text{HK}}$, which yields an expected rotation period close to the observed one \citep[using ][calibration]{Mamajek_2008} also allows us to rule out the possibility that the adopted rotation period is an alias of the true one.
From the analysis of the spectra (\Autoref{sec:stellar}), a marginal detection is obtained for Li $6707$ \angstrom doublet, with an EW measurement of $2.5 \pm0.2$ m\angstrom and an abundance (including NLTE effects) A(Li) $0.08 \pm 0.06$. In the $T_{\rm eff}-EW_{\rm Li}$ parameter space of the target, the lithium line is not a robust age diagnostic, and we can infer only a lower limit of $\sim 450$ Myr from \cite{jeffries2023}.
Finally, there are no indications of comoving objects in \textit{Gaia} DR3 and the star does not result member of known moving groups\footnote{Analysis with Banyan $\Sigma$ on-line tool \citep{banyan} yields a non-zero ($24\%$) membership probability to Upper Sco association. However, several stellar properties of K2-233 (lithium, age from isochrone, rotation and activity) are definitely not compatible with the very young age of the group, as also mentioned by \cite{David_2018}.}. The kinematic parameters are within the typical locus of young stars \citep{montes2001}.
From the combination of the various indicators, we adopt an age of $\tau_{\text{star}} = 400^{+200}_{-150}$ Myr.
This value is consistent with that found by \cite{David_2018},  supporting the classification of K2-233 as a moderately young stellar system.

To derive the stellar mass and radius of K2-233, we employed the web interface of the Bayesian inference tool \texttt{PARAM v1.3} \citep{DaSilva_2006}, which is based on \texttt{PARSEC} stellar isochrones \citep{Bressan_2012_parsec}.
The input quantities include $T_{\mathrm{eff}}$, [Fe/H], $\log g$, the magnitude in $V$ band and the parallax. These observables were used as Gaussian priors to compute the posterior probability distributions of the fundamental stellar parameters. Among the stellar parameters inferred by \texttt{PARAM} also shows the stellar age. Given the results obtained from the various indicators and since isochrone fitting is not able to put a strong constraint on the age of Main Sequence stars we decided to constrain $\tau_{\text{star}}$ between $200$ and $700$ Myr, following \cite{desidera2015}.
From this analysis, we obtained a stellar mass of $M_{\text{star}} = 0.795 \pm 0.009\:M_\odot$ and a radius of $R_{\text{star}} = 0.692 \pm 0.004 \:R_\odot$. These values are consistent with those found by \cite{David_2018} and \cite{2020lillo-box}. The errorbars do not include the systematic uncertainties on stellar models.

\section{Data analysis}
\label{sec:data-analysis}

We carried out a comprehensive photometric and spectroscopic analysis of the K2-233 system, aimed at characterising the transiting planets while mitigating the impact of the stellar activity. Our analysis is structured into three main steps: first, we modelled the K2 and TESS light curves to retrieve the planetary radii and transit parameters (\Autoref{subsec:lc_analysis}). Second, we performed a RV analysis exploring various Multi-GP frameworks – including the simultaneous modelling of photometry – to disentangle the stellar activity from the planetary signals and derive the masses (\Autoref{subsec:rv_analysis}). Finally, we investigated the system for additional planetary companions (\Autoref{subsec:another-planet}).
All analyses were conducted using the \texttt{PyORBIT} Python package \citep{Malavolta_2016, Malavolta_2018}, which includes the \texttt{dynesty} \citep{josh_speagle_2019} Nested Sampler algorithm, employed throughout this work. In each analysis, we employed 1500 live points in the nested sampler.
Nested samplers provide the Bayesian evidence of a model – indicated with $\mathcal{Z}$ – which allows us to perform a quantitative comparison between two different models applied to the same dataset: the greater the value, the better the model fits the data. Precisely – according to the scale of \cite{jeffreys1998theory} – a difference in $\log\mathcal{Z}$ greater than $2.5$ corresponds to moderate evidence, while a difference greater than $5.0$ indicates that the model with greater Bayesian evidence is highly preferred.
In this work, we systematically used these criteria to perform model selection. Most notably, for each configuration, we tested two orbital models to assert the best-fit scenario: one assuming circular orbits (with eccentricity fixed to zero) and one allowing for fully eccentric orbits with eccentricity as a free parameter for each planet. For all eccentric-orbit models, we adopted a Beta distribution prior for the eccentricity, with shape parameters $(a,b) = (1.52, 29)$, following the arguments \cite{Eylen_2019}.
Finally, for each dataset used in all the analyses, we adopted distinct jitter terms and offsets.

\subsection{Photometric analysis}
\label{subsec:lc_analysis}

For the data of K2-233 the native cadences of the K2 and TESS observations are 30 minutes and 2 minutes, respectively. To optimise the computational time, we re-binned the out-of-transit data from both instruments to 3-hours intervals without affecting the reliability of the time series to robustly describe the stellar signal.
Both the K2 and TESS light curves exhibit substantial variability induced by stellar activity. To model this modulation, for each dataset we implemented a GP with a QP kernel:

\begin{equation}
    \gamma_{\text{QP}} = h^2 \exp\Bigg\{
        - \frac{ \sin^2[ \pi(t_i - t_j) / P_{\text{rot}} ] }{ 2 \lambda_\text{P}^2 }
        - \frac{ (t_i - t_j)^2 }{ 2 \lambda_\text{e}^2 }
    \Bigg\}
\end{equation}

\noindent where the hyperparameters are the characteristic period $P_{\text{rot}}$ (corresponding to the stellar rotation period), the inverse harmonic complexity $\lambda_\text{P}$, the evolutionary timescale $\lambda_\text{e}$, and the correlation amplitude $h$.

Along with the GPs, we implemented simultaneously the transits model for each planet using the \texttt{batman} Python package \citep{Kreidberg_2015} using the quadratic law to describe the limb darkening \citep{Kopal_1950} with the parametrisation introduced by \cite{Kipping_2013}. For each planet, the parameters that describe a transit are the time of inferior conjunction $T_0$, the orbital period $P$, the eccentricity $e$, the argument of periastron $\omega$, the orbit inclination $i$, and the scaled planetary radius $R_{\text{p}}/R_{\text{star}}$. Another required parameter is the scaled semi-major axis $a/R_{\text{star}}$, but we preferred to use the stellar density $\rho_{\text{star}}$ as parameter to recover $a/R_{\text{star}}$ from the Kepler's third law (see \citealt{Winn_2010}), so that it is common for all the planets in the system.

We imposed that the two GPs used in the K2 and TESS light curves have to share the same $P_{\text{rot}}$, since it is not expected to change appreciably over the timespan that separates the two datasets. All the other hyperparameters were set as independent for each GP.
We used Gaussian priors centred on the values reported by \cite{David_2018} for the orbital periods and times of inferior conjunction, as well as a Gaussian prior for the stellar density based on the stellar mass and radius derived in \Autoref{subsec:age-mass-radius}. Uniform priors were used for all remaining parameters, except for the planets' eccentricities in the eccentric orbits model. 
The prior distributions and parameter bounds used in the nested sampling framework are listed in \Autoref{sec:table:priors}.

Assuming a circular-orbits configuration for the planets we found a stellar rotation period of $P_{\text{rot}} = 9.613^{+0.055}_{-0.059}$ days. 
This value is consistent with previous estimates in the literature.
\cite{David_2018} measured a rotation period of $9.754 \pm 0.038$ days using the autocorrelation function of the K2 photometry, while \cite{2020lillo-box} and \cite{Barrag_n_2023} obtained $9.638 \pm 0.083$ and $9.746^{+0.065}_{-0.070}$ days, respectively, from HARPS RV time series. The radii for the three planets retrieved in this model are equal to $R_b = 1.248 \pm 0.021$ $R_\oplus$, $R_c = 1.186 \pm 0.026$ $R_\oplus$ and $R_d = 2.301 \pm 0.035$ $R_\oplus$ for planets b, c, and d, respectively.
Although these radii are consistent within $3\sigma$ with previous determinations (\citealt{David_2018}; \citealt{2020lillo-box}; \citealt{Barrag_n_2023}), they are systematically smaller. This discrepancy may arise from the different treatment of stellar variability in the photometric analysis. Previous studies first removed the stellar modulation from the light curves and subsequently fitted the planetary transits, whereas in this work the stellar variability and planetary transits were modelled simultaneously.

We also analysed the light curves allowing for free eccentricity, obtaining slightly different radii for the planets: $R_b = 1.242^{+0.023}_{-0.021}$ $R_\oplus$, $R_c = 1.177 \pm 0.029$ $R_\oplus$, and $R_d = 2.259^{+0.055}_{-0.050}$ $R_\oplus$, for planets b, c, and d, respectively. 
The difference of the returned Bayesian evidence between the models with circular and eccentric orbits is $\Delta \log\mathcal{Z}_{\,\text{circ – ecc}} = -8$, suggesting that the best-fit model is that with eccentric orbits.
However, the distributions of the eccentricities of the three planets return values consistent with zero ($e_b = 0.063^{+0.038}_{-0.024}$; $e_c = 0.093^{+0.043}_{-0.039}$; $e_d = 0.038^{+0.032}_{-0.024}$) suggesting that the orbits of the three planets can be approximated as circular. 
Therefore, we chose as reference radii the values obtained in the circular-orbit configuration. 

We investigated the presence of transit timing variations (TTVs) for the three planets. Following the methodology described in \cite{Naponiello2026a}, we analysed the transits of each planet using a shared set of priors, while allowing individual transit times to vary. We find that the measured transit times are consistent with linear ephemerides, with deviations smaller than the average timing uncertainties. However, for K2-233 d only three transits are available, so this assessment should be considered preliminary.

\subsection{Radial velocity analysis}
\label{subsec:rv_analysis}

As shown in the GLS periodograms in \Autoref{sec:data:GLS}, the RV time series is dominated by the stellar activity signal. Similarly to the procedure applied in \Autoref{subsec:lc_analysis}, in order to model this signal in the data and retrieve those of the planets, we used a GP with a QP kernel along with a Keplerian function for each planet, which depends on the radial velocity semi-amplitude $K$, the eccentricity $e$, the argument of periastron $\omega$, the orbital period $P$, and the orbit's mean longitude $\lambda$ (which is strictly related to the time of inferior conjunction $T_0$).
For $P$ and $T_0$ we implemented Gaussian priors according from the results obtained in the photometry analysis (\Autoref{subsec:lc_analysis}).
As we performed with the light curve analyses, for the HARPS and HARPS-N dataset we used two different $\lambda_{\text{P}}$ and $\lambda_{\text{e}}$, while the $P_{\text{rot}}$ is shared among them.
This choice is justified by the large temporal gap between the two datasets (over three years), over which we expect a substantial variation of the spot configuration.
\cite{Barrag_n_2023} demonstrated that for K2-233 the Multi-GP approach provides more precise planetary signal recovery rather than the one-dimensional GP.
In the general case, the Multi-GPs consist in a certain number N of GP models relative to N time series. Indicating with $\mathcal{A}_i$ the $i$-th time series, the Multi-GP generate the following equation system:

\begin{equation}
    \label{sec:eq:multigp-general}
    \begin{aligned}
        \mathcal{A}_1 &=  A_1 \: G(t) + B_1 \: \dot{G}(t) \\
        \mathcal{A}_2 &=  A_2 \: G(t) + B_2 \: \dot{G}(t) \\
        \ldots \\
        \mathcal{A}_N &=  A_N \: G(t) + B_N \: \dot{G}(t) \\
    \end{aligned}
\end{equation}

\noindent where $G(t)$ and $\dot{G}(t)$ are a latent function obtained from the GP and its first derivative, while the coefficient $A_1,\,B_1,\,\ldots,\,A_N,\,B_N$ are the respective amplitudes and they are free parameters of the models.
As described in \cite{Rajpaul_2015}, photometric-like parameters – as the FWHM of the CCF and H$\alpha$ of the spectra – do not depend on $\dot{G}(t)$ which is related to the velocity of the active region on the stellar surface. For these parameters the amplitude $B_i$ is fixed to be zero.
For our analyses we set different values of the amplitudes $A_i$ and $B_i$ for the HARPS and HARPS-N datasets. 

Since K2-233 is a highly active star, we explored several Multi-GP configurations using different combinations of RVs and activity-indicator time series. To identify the most suitable tracers, we first modelled each activity indicator individually using a one-dimensional GP. For the \ion{Ca}{ii} time series, the GP hyperparameters were poorly constrained, with posterior distributions remaining largely unconstrained across the explored parameter space. This indicates that the stellar activity modulation is not robustly traced by this indicator in the available data. We therefore excluded \ion{Ca}{ii} from the subsequent analyses and retained only the BIS and FWHM of the CCF, together with the H$\alpha$ index.
We then performed five different three-dimensional Multi-GP analyses using different combinations of these activity indicators, as well as a four-dimensional Multi-GP including all of them.
In addition, TESS observed K2-233 during Sector 91, partially overlapping the HARPS-N observing campaign. We exploited this temporal overlap to perform a simultaneous modelling of stellar activity in both the photometric and spectroscopic datasets and to assess whether a joint analysis could improve the inferred planetary parameters.
Because the TESS photometry was collected contemporaneously with the HARPS-N observations, the activity signals present in the two datasets are expected to arise from the same stellar surface features. To account for this, we incorporated the TESS light curve into the Multi-GP framework adopted for the HARPS-N data, since the hyperparameters of the QP kernel are directly linked to the properties of active regions on the stellar surface. Since photometric observables are insensitive to the velocity of active regions, we set the coefficient associated with the $\dot{G}(t)$ term in \Eqref{sec:eq:multigp-general} to zero for the TESS light curve.
As our objective was to model the stellar activity contribution in the RV time series, the planetary transits were masked in the TESS light curve prior to the analysis. For each configuration, we therefore performed a joint analysis of the HARPS and HARPS-N RVs together with the TESS photometry, imposing a common stellar rotation period, $P_{\rm rot}$, across all datasets and adopting a single Multi-GP framework to jointly model the HARPS-N RVs and the TESS light curve.

\begin{table*}
    \centering
    \caption{List of the models used to analyse the HARPS and HARPS-N dataset in order to retrieve the planets' masses.}
    \label{tab:several-analyses}
    \small
    \begin{tabular*}{\textwidth}{@{\extracolsep{\fill}}cllcccccc}
        \toprule \toprule
        \textbf{N.} & \textbf{HARPS data used} & \textbf{HARPS-N data used} &
        \boldmath{$K_b$} \textbf{[m s$^{-1}$]} &  \boldmath{$K_c$} \textbf{[m s$^{-1}$]}  & \boldmath{$K_d$} \textbf{[m s$^{-1}$]} &
        \boldmath{$M_b$ [$M_\oplus$]} &  \boldmath{$M_c$ [$M_\oplus$]}  & \boldmath{$M_d$ [$M_\oplus$]} \\[3pt]
        \midrule
        1 & BIS + H$\alpha$ & BIS + H$\alpha$ &
        $1.57_{-0.61}^{+0.64}$ & 
        $2.82_{-0.44}^{+0.52}$ & 
        $1.30 \pm 0.42$ & 
        $2.8 \pm 1.2$ & 
        $7.2_{-1.1}^{+1.3}$ & 
        $5.1 \pm 1.6$  \\[4pt]
        2 & BIS + H$\alpha$ & BIS + H$\alpha$ + LC &
        $1.48^{+0.76}_{-0.69}$ & 
        $2.52 \pm 0.60$ & 
        $1.63^{+0.49}_{-0.51}$ & 
        $2.7 \pm 1.4$ & 
        $6.5 \pm 1.5$ & 
        $6.3 \pm 2.0$  \\[4pt]
        3 & BIS + FWHM & BIS + FWHM &
        $1.03^{+0.72}_{-0.61}$ & 
        $2.56 \pm 0.75$ & 
        $0.86^{+0.63}_{-0.53}$ & 
        $1.9^{+1.3}_{-1.1}$ & 
        $6.6 \pm 1.9$ & 
        $3.2^{+2.0}_{-2.5}$ \\[4pt]
        4 & BIS + FWHM  & BIS + FWHM + LC &
        $1.37^{+0.82}_{-0.74}$ & 
        $2.06 \pm 0.71$ & 
        $1.31^{+0.69}_{-0.65}$ & 
        $2.5^{+1.5}_{-1.3}$ & 
        $5.3 \pm 1.8$ & 
        $5.1^{+2.7}_{-2.5}$  \\[4pt]
        5 & FWHM + H$\alpha$ & FWHM + H$\alpha$ &
        $1.77^{+0.82}_{-0.79}$ & 
        $1.68^{+0.57}_{-0.53}$ & 
        $2.01 \pm 0.49$ & 
        $3.2 \pm 1.5$ & 
        $4.3 \pm 1.5$ & 
        $7.8 \pm 1.9$ \\[4pt]
        6 & FWHM + H$\alpha$ & FWHM + H$\alpha$ + LC &
        $2.14 \pm 1.00$ & 
        $1.63^{+0.60}_{-0.56}$ & 
        $2.01^{+0.48}_{-0.44}$ & 
        $3.9 \pm 1.9$ & 
        $4.2^{+1.6}_{-1.4}$ & 
        $7.8^{+1.9}_{-1.7}$ \\[4pt]
        7 & BIS + H$\alpha$ & FWHM + H$\alpha$ &
        $1.65^{+1.10}_{-0.97}$ & 
        $3.3 \pm 1.2$ & 
        $0.59^{+0.63}_{-0.52}$ & 
        $3.0^{+2.1}_{-1.8}$ & 
        $8.6 \pm 3.1$ & 
        $2.6^{+2.5}_{-1.6}$ \\[4pt]
        8 & BIS + H$\alpha$ & FWHM + H$\alpha$ + LC &
        $1.55^{+0.71}_{-0.69}$ & 
        $2.39^{+0.59}_{-0.56}$ & 
        $1.95 \pm 0.59$ &
        $2.8 \pm 1.3$ &
        $6.1 \pm 1.5$ & 
        $7.6 \pm 2.3$ \\[4pt]
        9 & FWHM+ H$\alpha$ & BIS + H$\alpha$ &
        $1.89^{+0.85}_{-0.78}$ & 
        $1.83^{+0.48}_{-0.50}$ & 
        $1.45^{+0.52}_{-0.48}$ & 
        $3.4 \pm 1.5$ & 
        $4.7 \pm 1.3$ & 
        $5.6 \pm 2.0$ \\[4pt]
        10 & FWHM + H$\alpha$ & BIS + H$\alpha$ + LC &
        $2.88^{+0.81}_{-0.83}$ & 
        $1.75^{+0.54}_{-0.58}$ & 
        $1.70 \pm 0.42$ & 
        $5.2 \pm 1.5$ &
        $4.5 \pm 1.5$ & 
        $6.6 \pm 1.6$ \\[4pt]
        11 & BIS + FWHM + H$\alpha$ & BIS + FWHM + H$\alpha$ &
        $1.23 \pm 0.57$ & 
        $2.06^{+0.54}_{-0.59}$ & 
        $1.38^{+0.47}_{-0.43}$ &
        $2.2 \pm 1.0$ & 
        $5.3 \pm 1.5$ & 
        $5.4 \pm 1.8$ \\[4pt]
        12 & BIS + FWHM + H$\alpha$ & BIS + FWHM + H$\alpha$ + LC &
        $1.8^{+1.2}_{-1.0}$ & 
        $2.3 \pm 1.2$ & 
        $1.46^{+0.87}_{-0.84}$ & 
        $3.3^{+2.2}_{-1.9}$ & 
        $5.9 \pm 3.1$ & 
        $5.7^{+3.4}_{-3.2}$ \\[4pt]
        \midrule \midrule
        \textbf{N.} & \textbf{HARPS data used} & \textbf{HARPS-N data used} &
        \multicolumn{2}{c}{\boldmath{$j_\textbf{HARPS}$} \textbf{[m s$^{-1}$]}} &  \multicolumn{2}{c}{\boldmath{$j_\textbf{HARPS-N}$} \textbf{[m s$^{-1}$]}} &
        \multicolumn{2}{c}{\textbf{MGIC + 1000}} \\[3pt]
        \midrule
        1 & BIS + H$\alpha$ & BIS + H$\alpha$ & 
        \multicolumn{2}{c}{$1.59 \pm 0.47$} & 
        \multicolumn{2}{c}{$5.43^{+0.89}_{-0.75}$} &
        \multicolumn{2}{c}{$-184$} \\[4pt]
        2 & BIS + H$\alpha$ & BIS + H$\alpha$ + LC &
        \multicolumn{2}{c}{$1.53 \pm 0.51$} &
        \multicolumn{2}{c}{$2.7^{+2.3}_{-1.8}$} &
        \multicolumn{2}{c}{$-99$} \\[4pt]
        3 & BIS + FWHM & BIS + FWHM &
        \multicolumn{2}{c}{$4.74^{+0.70}_{-0.72}$} & 
        \multicolumn{2}{c}{$6.0 \pm 1.4$} & 
        \multicolumn{2}{c}{$-160$} \\[4pt]
        4 & BIS + FWHM & BIS + FWHM + LC &
        \multicolumn{2}{c}{$4.57^{+0.73}_{-0.80}$} & 
        \multicolumn{2}{c}{$1.7^{+1.5}_{-1.2}$} &
        \multicolumn{2}{c}{$-70$} \\[4pt]
        5 & FWHM + H$\alpha$ & FWHM + H$\alpha$ &
        \multicolumn{2}{c}{$4.15 \pm 0.38$} & 
        \multicolumn{2}{c}{$9.9^{+1.7}_{-1.3}$} & 
        \multicolumn{2}{c}{$-151$} \\[4pt]
        6 & FWHM + H$\alpha$ & FWHM + H$\alpha$ + LC &
        \multicolumn{2}{c}{$2.92^{+0.45}_{-0.47}$} & 
        \multicolumn{2}{c}{$8.6^{+1.6}_{-1.8}$} &
        \multicolumn{2}{c}{$-79$} \\[4pt]
        7 & BIS + H$\alpha$ & FWHM + H$\alpha$ &
        \multicolumn{2}{c}{$17.8 \pm 1.1$} &
        \multicolumn{2}{c}{$7.3 \pm 1.2$} &
        \multicolumn{2}{c}{$59$} \\[4pt]
        8 & BIS + H$\alpha$ & FWHM + H$\alpha$ + LC &
        \multicolumn{2}{c}{$2.65^{+0.42}_{-0.39}$} &
        \multicolumn{2}{c}{$33.4^{+6.4}_{-4.7}$} &
        \multicolumn{2}{c}{$-70$} \\[4pt]
        9 & FWHM+ H$\alpha$ & BIS + H$\alpha$ &
        \multicolumn{2}{c}{$2.81^{+0.40}_{-0.37}$} &
        \multicolumn{2}{c}{$7.76^{+1.00}_{-0.86}$} &
        \multicolumn{2}{c}{$-171$} \\[4pt]
        10 & FWHM + H$\alpha$ & BIS + H$\alpha$ + LC &
        \multicolumn{2}{c}{$1.91 \pm 0.40$} & 
        \multicolumn{2}{c}{$2.4 \pm 1.3$} &
        \multicolumn{2}{c}{$-88$} \\[4pt]
        11 & BIS + FWHM + H$\alpha$ & BIS + FWHM + H$\alpha$ &
        \multicolumn{2}{c}{$3.89 \pm 0.45$} & 
        \multicolumn{2}{c}{$7.29^{+0.72}_{-0.74}$} &
        \multicolumn{2}{c}{$-173$} \\[4pt]
        12 & BIS + FWHM + H$\alpha$ & BIS + FWHM + H$\alpha$ + LC & 
        \multicolumn{2}{c}{$18.18 \pm 0.94$} & 
        \multicolumn{2}{c}{$24.5^{+2.1}_{-1.8}$} &
        \multicolumn{2}{c}{$163$} \\[4pt]
        \bottomrule
    \end{tabular*}
    \tablefoot{Each row represent a model in which the activity indicators of HARPS and HARPS-N data used along with the respective RVs for the analysis are indicated in second and third column, respectively.
    Other columns indicate the radial velocity semi-amplitudes and respective masses of the planets of K2-233 retrieved by the different analyses, along with the corresponding jitters of both HARPS and HARPS-N RV time series, and the MGIC factor in logarithm units increased by 1000 for better readability. 
    Uncertainties correspond to the 16th and 84th percentiles of the posterior distributions.}
\end{table*}

\begin{figure*}
\centering
    \includegraphics[width=\textwidth]{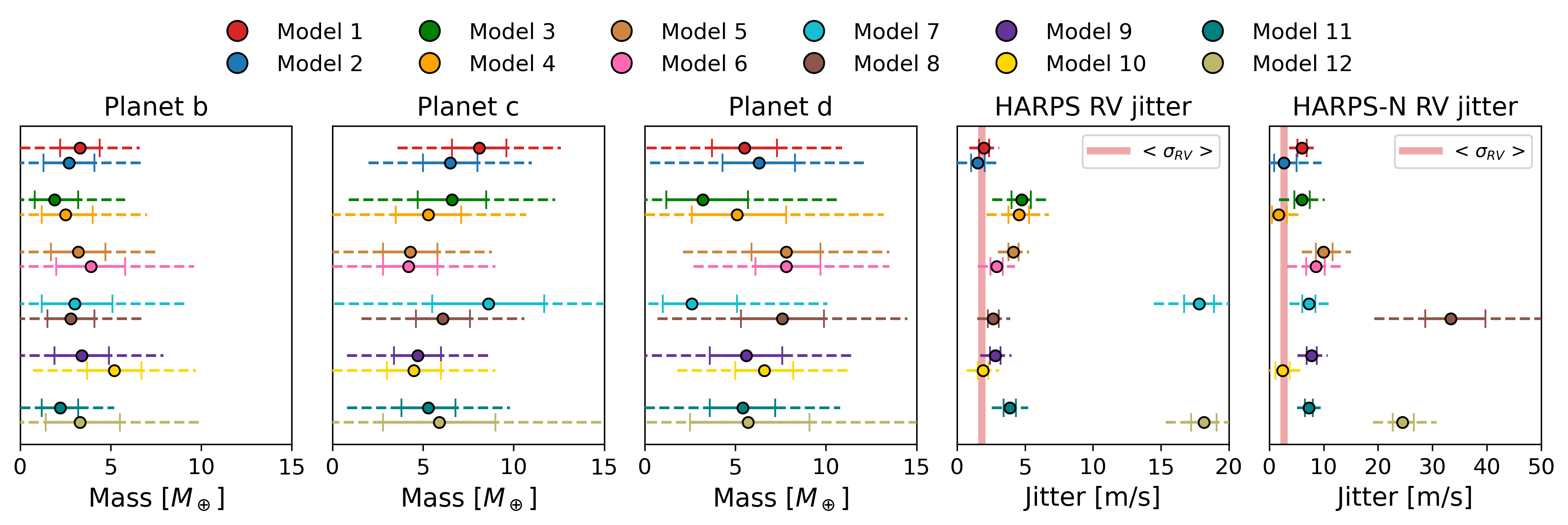}
    \caption{Starting from left, masses values of planet b, c, and d, and jitters of HARPS and HARPS-N RV time series obtained from the analyses using different combinations of RV, BIS and FWHM of the CCF, H$\alpha$ of the spectra, and the TESS light curve.
    Solid lines correspond to uncertainties to the 16th and 84th percentiles of the posterior distributions, while dashed lines correspond to $0.1$th and $99.9$th percentile. Vertical solid lines in the jitters' panels indicate the corresponding median RV uncertainty.}
    \label{sec:data:analyses-comparison}
\end{figure*}

\begin{figure*}
\centering
    \includegraphics[width=.8\textwidth]{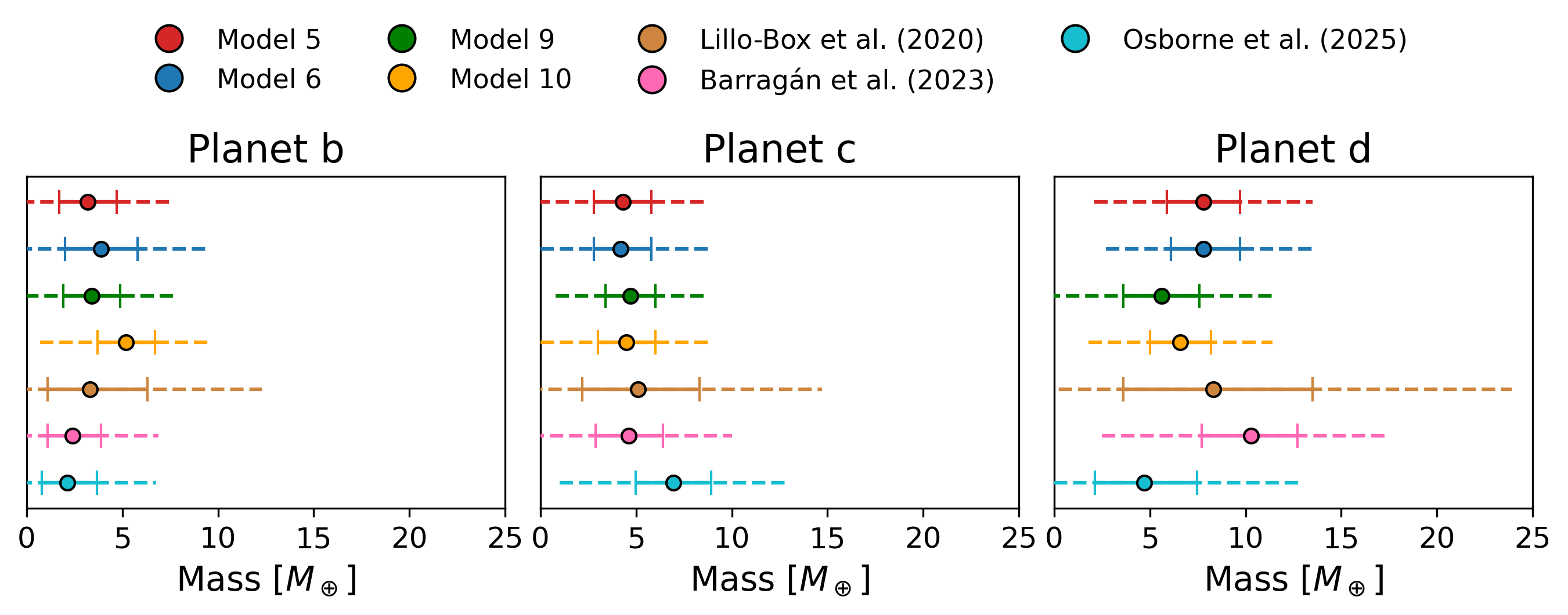}
    \caption{Starting from left, masses values of planet b, c, and d obtained from this work (in red, blue, green, and orange), \cite{2020lillo-box} (in brown), \cite{Barrag_n_2023} (in pink), and \cite{Osborne_2024} (in cyan). 
    Uncertainties correspond to the 16th and 84th percentiles of the posterior distributions.
    Solid lines correspond to uncertainties to the 16th and 84th percentiles of the posterior distributions, while dashed lines correspond to $0.1$th and $99.9$th percentile.}
    \label{sec:data:mass-comparison}
\end{figure*}

\begin{table*}
    \centering
    \caption{Parameters of the planets of K2-233 obtained from the joint light curve and RV analysis in both circular and eccentric orbits configuration using the Model 9 of \Autoref{tab:several-analyses}.}
    \label{tab:results-planets}
    \small
    \begin{tabular*}{\textwidth}{@{\extracolsep{\fill}}lcccccc}
        \toprule \toprule
        \textbf{Parameter} & \multicolumn{2}{c}{\textbf{Planet b}} & \multicolumn{2}{c}{\textbf{Planet c}} & \multicolumn{2}{c}{\textbf{Planet d}} \\[3pt]
        & Circular orbit & Eccentric orbit & Circular orbit & Eccentric orbit & Circular orbit & Eccentric orbit \\
        \midrule
        $K$ [m s$^{-1}$] & $1.89^{+0.85}_{-0.78}$ & $2.18 \pm 0.68$ & $1.83^{+0.50}_{-0.48}$ & $2.03^{+0.48}_{-0.42}$ & $1.45^{+0.52}_{-0.48}$ & $1.63^{+0.57}_{-0.52}$ \\[4pt]
        $M_p$ [$M_\oplus$] & $3.4 \pm 1.5$ & $3.9 \pm 1.2$ & $4.7 \pm 1.3$ & $5.2 \pm 1.2$ & $5.6 \pm 2.0$ & $6.3^{+2.2}_{-2.0}$ \\[4pt]
        $R_p$ [$R_\oplus$] & $1.248 \pm 0.021$ & $1.242^{+0.023}_{-0.021}$ & $1.186 \pm 0.026$ & $1.177 \pm 0.029$ & $2.301 \pm 0.035$ & $2.259^{+0.055}_{-0.050}$ \\[4pt]
        $\rho_p$ [$\rho_\oplus$] & $1.74 \pm 0.78$ & $2.01 \pm 0.62$ & $2.82 \pm 0.78$ & $3.11 \pm 0.74$ & $0.46 \pm 0.17$ & $0.52^{+0.18}_{-0.16}$ \\[4pt]
        $a$ [AU] & $0.03309 \pm 0.00013$ & $0.03309 \pm 0.00013$ & $0.06669 \pm 0.00025$ & $0.06669 \pm 0.00025$ & $0.15231 \pm 0.00058$ & $0.15231 \pm 0.00058$  \\[4pt]
        $e$ & $0$ & $0.074^{+0.062}_{-0.025}$ & $0$ & $0.064^{+0.034}_{-0.023}$ & $0$ & $0.012 \pm 0.007$ \\[4pt]
        $i$ [deg] & $89.45^{+0.39}_{-0.51}$ & $89.31^{+0.47}_{-0.62}$ & $89.77^{+0.16}_{-0.22}$ & $89.68^{+0.23}_{-0.32}$ & $89.48 \pm 0.05$ & $89.62^{+0.20}_{-0.13}$ \\[4pt]
        $T_{\text{eq}}$ [K] & $1568 \pm 19$ & $1544 \pm 22$ & $1104 \pm 14$ & $1087 \pm 15$ & $731 \pm 9$ & $720 \pm 10$ \\[4pt]
        $\tau_{14}$ [hours] & $1.969^{+0.035}_{-0.048}$ & $1.904^{+0.050}_{-0.071}$ & $2.792^{+0.048}_{-0.059}$ & $2.699^{+0.070}_{-0.102}$ & $3.942^{+0.129}_{-0.123}$ & $3.972^{+0.232}_{-0.094}$ \\[4pt]
        \bottomrule
    \end{tabular*}
    \tablefoot{Uncertainties correspond to the 16th and 84th percentiles of the posterior distributions. Starting from the top the parameters indicate the RV semi-amplitude, the planet's mass, the planet's radius, the planet's bulk density, the semi-major axis, the eccentricity, the orbit's inclination, the planet's equilibrium temperature, and the transit duration. Others parameters are shown in \Autoref{sec:table:priors}.}
\end{table*}

In \Autoref{tab:several-analyses} and \Autoref{sec:data:analyses-comparison} we present the RV semi-amplitudes and corresponding masses obtained for the three planets. In all cases, the inferred mass of planet c implies a high bulk density, ranging from $2.54$ $\rho_\oplus$ to $4.36$ $\rho_\oplus$. To assess which solutions are physically plausible, we considered the mass--radius relation for pure-iron planets derived by \cite{Fortney2007}. For the measured radius of planet c ($R_c = 1.186 \pm 0.026$ $R_\oplus$), this relation predicts a mass of $M_{c,\,\mathrm{Fe}} = 5.0 \pm 0.5$ $M_\oplus$. We therefore regard significantly larger values as physically unlikely.
We emphasise that this criterion strongly depends on the adopted planetary radius, which may itself be affected by stellar activity. In particular, for active stars such as K2-233, starspots can alter the observed transit depth and therefore bias the inferred planetary radius \citep{Szabo2021}. To assess the possible impact of stellar activity on the radius of planet c, following \cite{Szabo2021} we independently fitted the individual transits. Transits yielding robust radius estimates range from $1.01$ to $1.43\ R_\oplus$. The dispersion among the individual measurements exceeds their typical formal uncertainties, indicating additional transit-to-transit scatter that may arise from stellar activity, residual systematics, or variations in data quality. We further examined the individual radius estimates as a function of stellar rotational phase, but found no clear modulation or systematic trend. Thus, while we find no evidence for a rotationally modulated bias, the observed scatter does not allow us to completely exclude activity-related effects.
Nevertheless we retain the radius obtained in \Autoref{subsec:lc_analysis}, $R_c = 1.186 \pm 0.026\ R_\oplus$, as our reference value. Under the pure-iron mass–radius relation of \cite{Fortney2007}, this corresponds to $M_{c,\mathrm{Fe}} = 5.0 \pm 0.5\ M_\oplus$, which we adopt as a reference upper bound.
Under this criterion, only Models 5, 6, 9, and 10 yield physically plausible solutions. 
However, Models 5, 6, and 9 return HARPS-N jitter values substantially larger than the median RV uncertainty, suggesting that one or more components of the signal have not been adequately modelled and that residual variability remains in the data.
Model 10, on the other hand, yields a mass of $M_b = 5.2 \pm 1.5$ $M_\oplus$ for planet b, larger than in the other configurations. Nevertheless, this value remains physically plausible: using the pure-iron mass–radius relation of \cite{Fortney2007} and the measured radius of planet b ($R_b = 1.248 \pm 0.021\ R_\oplus$), the corresponding theoretical upper limit is $M_{b,\,\mathrm{Fe}} = 6.3 \pm 0.5$ $M_\oplus$.

\begin{figure*}
\centering
    \includegraphics[width=0.95\textwidth]{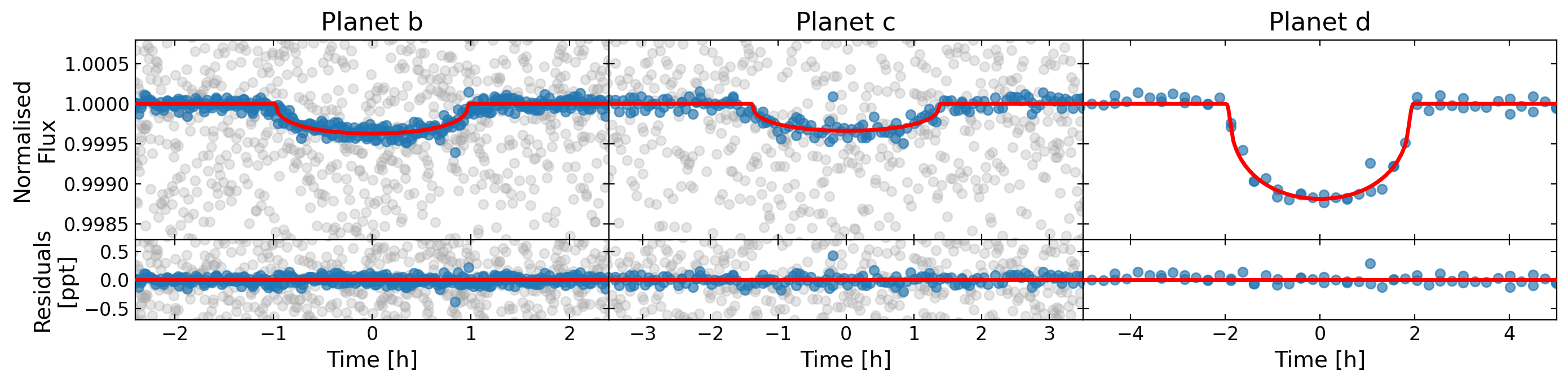}
    \caption{Top panels show the normalised phase-folded planetary transits of K2-233's planets from the light curves of K2 and TESS. Blue dots are 30 minutes re-binned data for a better readability, while grey dots represent the original cadence of the light curves. Red solid lines show the \texttt{batman} transit models obtained from the parameters retrieved from the nested sampling in \Autoref{subsec:lc_analysis}. Bottom panels show the residuals between the models and the respective normalised phase-folded planetary transits.}
    \label{sec:data:LC-fold}
\end{figure*}

\begin{figure*}
\centering
    \includegraphics[width=0.95\textwidth]{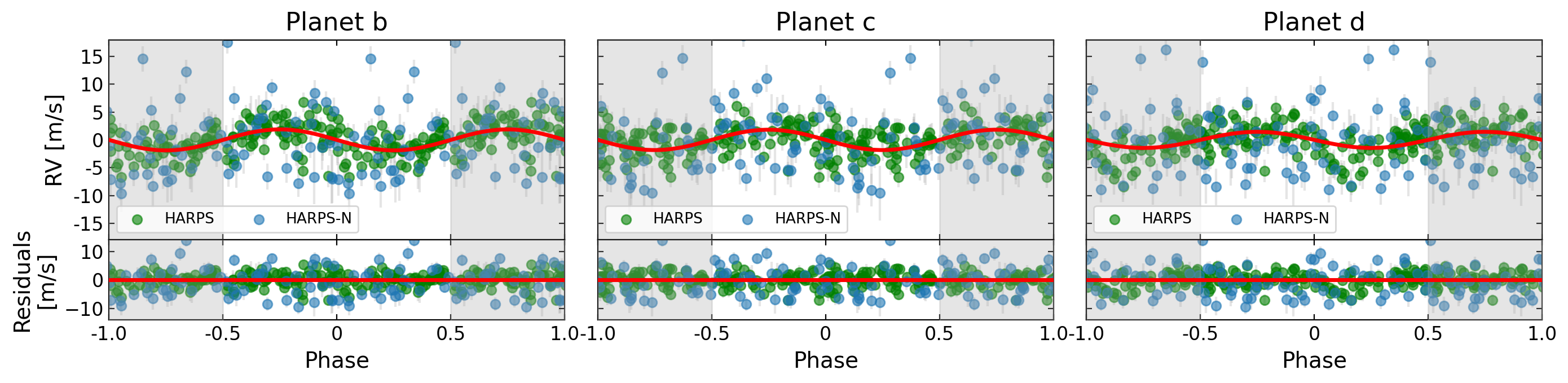}
    \caption{Top panels show the phase-folded RV fit of K2-233's planets. For each planet, the dots in each panel were obtained from the RV time series deprived from the contributions of the stellar activity and Keplerian signals from the other two planets obtained from the model 10 of \Autoref{tab:several-analyses}. Green and blue dots are referred to the RV from HARPS and HARPS-N, respectively.
    The grey-shaded regions represent phase extensions of the central (white) interval, where the data points are duplicated according to their orbital phase for a better readability.
    Red solid lines represent the corresponding Keplerian curve obtained from the parameters' values retrieved from the NS. Bottom panels show the residuals between the models and the phase-folded RV curve.}
    \label{sec:data:RV-fold}
\end{figure*}

To identify a preferred model, we adopted the statistical approach proposed by \cite{Barragan2026} for Multi-GP frameworks. The authors introduced the Multi-GP Information Criterion (MGIC), a statistical metric that combines the conditional RV likelihood with an effective parameter count accounting for the regularisation imposed by the Multi-GP model on the RV component. Lower MGIC values indicate a more favourable model.
We calculated the MGIC for all Multi-GP configurations considered in the analysis of K2-233, separately for the HARPS and HARPS-N datasets, and then computed the total MGIC as the sum of the two values. The results are reported in \Autoref{tab:several-analyses}. The lowest MGIC is obtained for Model 1, which therefore provides the statistically preferred description of the RV time series. However, the corresponding mass of planet c exceeds the pure-iron reference value of $M_{c,\,\mathrm{Fe}} = 5.0 \pm 0.5\ M_\oplus$ and is therefore difficult to reconcile with a physically plausible planetary composition. 
The MGIC provides a statistical criterion for comparing the relative performance of the different models, but it does not by itself enforce physical plausibility constraints. These additional considerations must therefore be assessed when interpreting the statistical results. 
Among the models yielding physically plausible masses, Model 9 has the lowest MGIC.
Although this model provides physically plausible planetary masses and lowest MGIC, we do not identify it as the uniquely preferred solution. The strong activity level of K2-233 makes the inferred planetary masses particularly sensitive to the adopted activity-modelling strategy, as demonstrated by their dependence on the selected activity indicators. Consequently, we consider all solutions consistent with the physical limits imposed by the \cite{Fortney2007} relation to be viable, namely Models 5, 6, 9, and 10.
Nevertheless, Model 9 was adopted for the subsequent analyses presented in this work.

In \Autoref{sec:data:mass-comparison} we compare the masses obtained from Models 5, 6, 9, and 10 with those reported by \cite{2020lillo-box}, \cite{Barrag_n_2023}, and \cite{Osborne_2024}. All of our results are consistent within $1\sigma$ with the previous estimates. However, Model 10 yields a slightly larger mass for planet b, while all four models return lower masses for planet d than those reported by \cite{Barrag_n_2023}.
We emphasise that \cite{Barrag_n_2023} derived a mass of $10.3^{+2.4}_{-2.6}$ $M_\oplus$ for planet d. This value exceeds the highest mass obtained in any of our analyses, namely $7.9 \pm 1.9$ $M_\oplus$ from Model 5. Such differences are likely a consequence of the different treatments of stellar activity adopted in the RV modelling and further support the conclusion that the planetary masses inferred for the K2-233 system are highly sensitive to the methodology used to model stellar activity.

All the models listed in \Autoref{tab:several-analyses} have circular-orbits configuration for all the three planets.
To investigate possible presence of eccentric orbits we re-run Model 9 setting the eccentricities as free parameters for all the planets. The retrieved values are $e_b = 0.055^{+0.041}_{-0.024}$, $e_c = 0.032^{+0.019}_{-0.016}$, and $e_d = 0.016^{+0.024}_{-0.010}$. These values are in agreement with zero within $3\sigma$, therefore they must interpreted as upper limits. The Bayesian Evidence difference between the models with circular- and eccentric-orbits configurations is $\Delta \log \mathcal{Z}_{\text{circ} - \text{ecc}} = 6$, indicating strong support for circular orbits, in contrast to the results obtained in \Autoref{subsec:lc_analysis}.
Although the Bayesian evidence clearly favours the circular-orbit solution, these eccentricity estimates may be interpreted as upper limits.
This outcome is also supported by theoretical considerations: in young, compact multi-planet systems without massive companions, such as K2-233, significant eccentricities are not generally expected. Planet–disc interactions during formation efficiently damp orbital eccentricities \citep{Kley_2012}, resulting in nearly circular architectures. The eccentricity constraints derived from the Keplerian fits are therefore consistent with expectations for systems of this type (e.g. \citealt{Eylen_2015}; \citealt{Eylen_2019}).
\Autoref{tab:results-planets} summarises the planetary parameters inferred from the nested sampling analysis for both the circular and eccentric orbits models applied to the Model 9 of \Autoref{tab:several-analyses}. The RV semi-amplitudes derived under the two assumptions agree within $1\sigma$, however, the median mass inferred for planet c in the eccentric-orbit configuration slightly exceeds the $M_{c,\,\mathrm{Fe}} = 5.0 \pm 0.5\ M_\oplus$ expected for a pure-iron composition according to the mass–radius relation of \cite{Fortney2007}. 
The posterior distributions of the planetary RV semi-amplitudes for the circular-orbit model are shown in \Autoref{sec:data:corners}, while \Autoref{sec:data:rv-gp-model} shows the RV time series together with the best-fitting GP model and the Keplerian signals of the three planets. Finally, \Autoref{sec:data:LC-fold} displays the corresponding folded transit light curves, while \Autoref{sec:data:RV-fold} presents the RV signals induced by each planet.

\subsection{Search for additional planetary companions}
\label{subsec:another-planet}

Photometric observations reported by \cite{David_2018} revealed three transiting planets orbiting K2-233. Nevertheless, additional non-transiting companions may still be present in the system.
To investigate this possibility, we re-analysed the RV time series using the same Model 9 described in \Autoref{subsec:rv_analysis}, but including an additional Keplerian signal with its orbital period constrained between 25 and 500 days.
The nested sampling analysis returned a candidate signal with an orbital period of $P_e = 449^{+35}_{-9}$ days and an RV semi-amplitude of $K_e = 1.99^{+0.53}_{-0.56}\ \text{m s}^{-1}$. 
The corresponding mass for this potential planet is $M_e \sin i = 15.7^{+7.1}_{-6.1}\ M_\oplus$.
However, the difference in Bayesian evidence between the three-planet model (i.e. that adopted in \Autoref{subsec:rv_analysis}) and the four-planets configuration is $\Delta \log\mathcal{Z}_{\,\text{3pl – 4pl}} = 15$, indicating that the model including only the three known transiting planets provides a significantly better description of the data.
To assess the RV time series detection completeness we performed an injection and retrieval scheme in which synthetic companions are injected into the post-fit residuals of the three-planets configuration and they are subjected to statistical tests to assess their detectability (see \Autoref{appendix:detection_compl} for details on this analysis).
From the results obtained from this analysis, we conclude that, if an additional planet is present in the system, the currently available spectroscopic observations lack the precision and/or temporal baseline required for its detection.

We also inspected the astrometric information available in the \textit{Gaia} DR3 archive \citep{gaia}, in particular the Renormalised Unit Weight Error (RUWE; \citealt{Lindegren_2018_ruwe}). The reported value for K2-233 is $0.929$, consistent with expectations for a well-behaved single-star astrometric solution and with the absence of multiplicity discussed in \Autoref{subsec:age-mass-radius}. Values significantly larger than unity (typically $\gtrsim 1.4$) are often interpreted as indicators of unresolved binarity or excess astrometric noise (e.g. \citealt{Stassun_2021}), but no such anomaly is observed in this case. 
We note, however, that a RUWE value close to unity does not exclude the presence of low-mass or wide-separation companions whose astrometric signal may fall below the current \textit{Gaia} sensitivity.
As an additional diagnostic, we considered the \texttt{IPD\_frac\_multi\_peak} parameter \citep{gaia_2}, which traces the fraction of image parameter determinations with multiple peaks and is sensitive to wide or partially resolved companions. This parameter ranges from $0$ to $100$, with values $\lesssim 2$ typically associated with single stars. For K2-233, \texttt{IPD\_frac\_multi\_peak} is equal to $0$, further supporting the absence of nearby companions.

\section{The role of photoevaporation}
\label{sec:photoevaporation}

Using XMM-Newton \citep{Jensen_2001_xmmnewton} observations, we estimated the X-ray flux incident on the three planets of the K2-233 system. These measurements were then used to investigate the impact of stellar XUV irradiation on the atmosphere of K2-233 d, which is the only planet of the system with a density consistent with the presence of substantial gaseous envelope. In particular, we modelled the atmospheric evolution of the planet, simulating its response to photoevaporative mass loss over the system’s lifetime, from $10$ Myr to $5$ Gyr.

\subsection{XMM-Newton's data analysis}
\label{sec:x-ray}

\begin{table}
\centering
\caption{Best fitting parameters from our X-ray analysis of K2-233.}
\label{tab:xray}
\begin{threeparttable}
\begin{tabular}{lcc}
\toprule
Parameter                                                  & Value                                             & Unit                   \\ \midrule
Temperature 1, $kT_1$                                      & $0.319^{+0.024}_{-0.027}$                         & keV                    \\
Emission Measure 1                                         & $1.08^{+0.11}_{-0.14}\times10^{51}$               & cm$^{-3}$              \\
Temperature 2, $kT_2$                                      & $0.99\pm0.14$                                     & keV                    \\
Emission Measure 2                                         & $\left(3.44^{+0.86}_{-0.76}\right)\times10^{50}$  & cm$^{-3}$              \\
Flux at Earth$^\dagger$                          & $\left(4.47^{+0.27}_{-0.23}\right)\times10^{-14}$ & erg\,s$^{-1}$\,cm$^{-2}$ \\
Flux at Earth$^\ddagger$                         & $\left(5.63^{+0.34}_{-0.31}\right)\times10^{-14}$ & erg\,s$^{-1}$\,cm$^{-2}$ \\
X-ray luminosity, $L_{\rm X}$ $^\ddagger$                   & $\left(3.09^{+0.19}_{-0.17}\right)\times10^{28}$  & erg\,s$^{-1}$           \\
 $L_{\rm X}^\ddagger/L_{\rm bol}$ & $\left(3.13^{+0.25}_{-0.23}\right)\times10^{-5}$  & dimensionless          \\
X-ray flux at K2-233b$^\ddagger$                            & $10000^{+610}_{-550}$                             & erg\,s$^{-1}$\,cm$^{-2}$ \\
X-ray flux at K2-233c$^\ddagger$                            & $2460^{+150}_{-140}$                              & erg\,s$^{-1}$\,cm$^{-2}$ \\
X-ray flux at K2-233d$^\ddagger$                            & $472^{+29}_{-26}$                                 & erg\,s$^{-1}$\,cm$^{-2}$ \\ \bottomrule
\end{tabular}
\tablefoot{Notes on energy bands: $^\dagger$ 0.3--2.4\,keV; $^\ddagger$ 0.1--2.4\,keV.}
\end{threeparttable}
\end{table}

We analysed the X-ray observations obtained with XMM-Newton and described in \Autoref{subsec:xmm-newton-data}.
Visual inspection of our extracted X-ray light curves revealed no obvious variation of the source within the uncertainties. We thus treated all good time intervals as representative of the quiescent activity.
We fitted the spectra from each EPIC camera simultaneously using \textsc{Xspec} 12.14.0h \citep{Arnaud1996}. With the spectra binned to a minimum of 10 counts per bin, we adopted C-statistics for finding the best fit \citep{Cash1979}. We used a two temperature APEC model \citep{Smith2001}, forcing the temperatures and normalisations to be the same across all three spectra, and fixing the coronal abundances to Solar values \citep{Asplund2009}. We also included a TBABS model for interstellar absorption \citep{Wilms2000}, and fixed the hydrogen column density along the line of sight to $2.4\times10^{18}$\,cm$^{-2}$, estimated using the LISM code \citep{Youngblood2025}.

We present the best fitting spectral model parameters from our X-ray analysis in \Autoref{tab:xray}. This table also provides fluxes for the ``observed" 0.3--2.4\,keV band, and the flux for the commonly used ROSAT band, 0.1--2.4\,keV, based on extrapolation of our best fitting model below 0.3\,keV. We also give X-ray luminosity and fluxes at each of the K2-233 planets in this latter band.

\subsection{K2-233 d atmospheric evolution}
\label{subsec:planet-d-photoevaporation}

Only the outermost planet (K2-233 d) exhibits a density consistent with the retention of a substantial gaseous envelope at its current age. To assess the current atmospheric escape, the past history and the fate of the planetary atmosphere under photo-evaporation we followed a modelling approach proposed in \cite{Locci19} and updated in other works \citep[e.g.][]{benatti2021, maggio2022, damasso2024, mantovan2024, nardiello2025}. In brief, this method relies on coupling MESA Stellar Tracks (MIST, \citealt{choi+2016}) and the temporal XUV luminosity evolution derived by \cite{Penz08a} and \cite{SF2025} with the planetary core-envelope models of \cite{Fortney2007} and \cite{Lopez_2014}, as well as the ATES photoionization hydrodynamics code \citep{Caldiroli+2021, Caldiroli+2022, spinelli2023}. 

By simulating a grid of exoplanets across varying levels of XUV irradiation and gravitational potential, \cite{Caldiroli+2022} derived an analytical approximation for the evaporation efficiency ($\eta_{\text{eff}}$), expressed as a function of the planetary gravitational potential ($\Psi$) and XUV irradiation. By incorporating the evaporation efficiency into the energy-limited framework of \cite{erkaev2007}, they derived an approximation for the atmospheric mass-loss rate ($\dot{M}$) consistent with hydrodynamic simulations:

\begin{equation}
\dot{M} = \eta_{\text{eff}}(F_{\text{XUV}}, \Psi) \frac{3 F_{\text{XUV}}}{4 G K \rho_p}
\end{equation}

In this expression, $F_{\text{XUV}}$ denotes the XUV flux at the orbital distance, $\rho_p$ is the mean planetary density, and the factor $K$ accounts for stellar tidal forces. This analytical approach was adopted to expedite the modelling of the planet's long-term evolution. 

To characterise the host star, we utilised the MESA Isochrones and Stellar Tracks \citep{choi+2016} to retrieve the stellar evolutionary tracks, specifically the bolometric luminosity ($L_{\text{bol}}$). We selected a track with  ($[\text{Fe/H}] = -0.08$) and a stellar mass of $0.769 \, M_{\odot}$, which is consistent within $1\sigma$ uncertainties with the star's current age, $L_{\text{bol}}$, effective temperature ($T_{\text{eff}}$), and metallicity.

\begin{figure}
    \centering
    \includegraphics[width=1.0\linewidth]{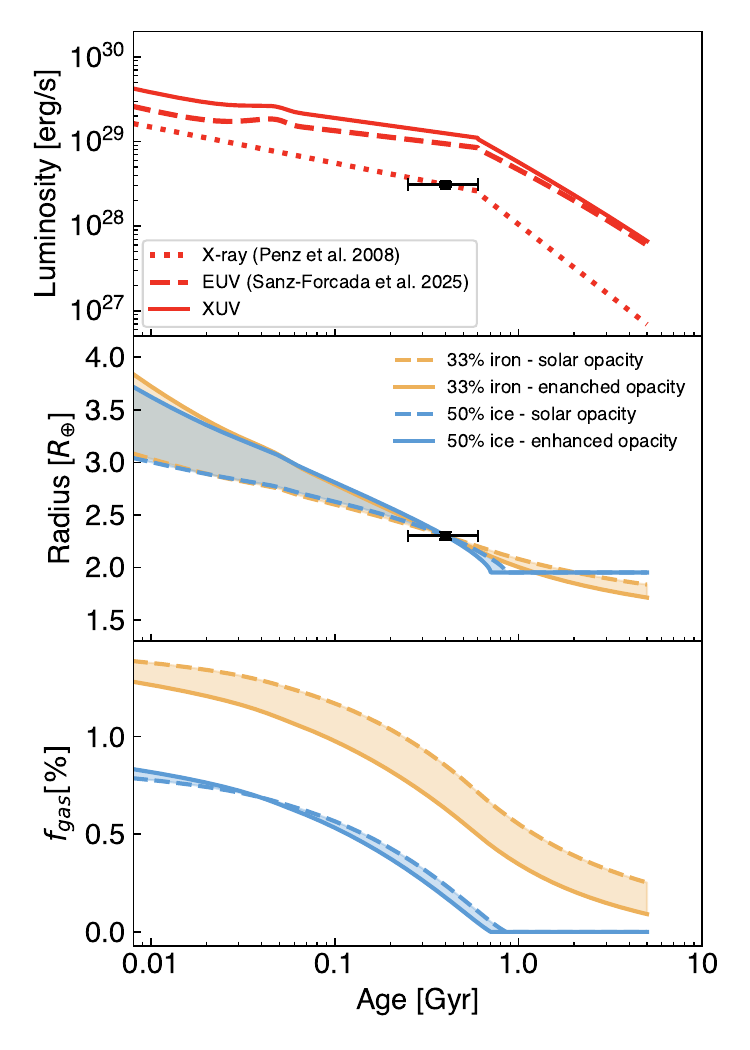}
    \caption{Top panel: time evolution of X-ray (5–100 Å), EUV (100–920 Å) and total XUV luminosity of K2-233 according to \cite{Penz08a} and the scaling relation by \cite{SF2025}. Uncertainties on the age and the measured X-ray luminosity of K2-233 are also indicated. Central panel: Evolution of planetary radius for K2-233 d according to our models with different assumptions. Black squares mark the current location of the planet. Bottom panel: evolution of the atmospheric mass fraction for K2-233 d according to our models with different assumptions.}
    \label{fig:xuv_radius_fenv}
\end{figure}

For the X-ray luminosity evolution we assume the X-ray luminosity versus age analytical relation derived by \cite{Penz08a}, anchored to the current value of the X-ray luminosity (\Autoref{sec:x-ray}). Then we estimate the stellar irradiation in the extreme-ultraviolet (EUV) band through the scaling law between EUV (100–920 Å), X-ray (5–100 Å) and bolometric luminosities derived by \cite{SF2025} (\Autoref{fig:xuv_radius_fenv}, top panel).

To model the photoevaporative evolution of the planetary atmosphere, we constrained the current core mass, atmospheric mass fraction ($f_{\text{gas}}$), and envelope radius using the core-envelope framework from \cite{Lopez_2014}. This model treats the planet as a two-component system defined by four variables: the core mass and radius ($M_{\text{core}}$, $R_{\text{core}}$), the envelope radius ($R_{\text{gas}}$), and the atmospheric mass fraction ($f_{\text{gas}}$). These parameters are solved by coupling two observational constraints - the total measured radius ($R_p$) and mass ($M_p$) - with the theoretical relations of \cite{Lopez_2014} that link $R_{\text{gas}}$, $f_{\text{gas}}$, and $M_p$ at the system's current age and orbital separation. For planetary mass we assume the internal structure models by \cite{Fortney2007} where cores can be composed of different ice-rock or rock-iron mixtures. 

\cite{Lopez_2014} relation was specifically calibrated for H/He dominated atmospheres. This framework incorporates the thermal evolution of the envelope, accounting for both cooling and contraction over time \cite{lopez2012}. Within this model, the temporal evolution of the radius is intrinsically tied to metallicity: for solar opacity, the envelope radius scales as $t^{-0.11}$, whereas enhanced opacity leads to a steeper decline of $t^{-0.18}$. Furthermore, the envelope size is subject to additional time-dependent variations when accounting for changes in bolometric luminosity and equilibrium surface temperature.

Assuming an Earth-like interior comprising 67\% rock and 33\% iron, we modelled the structure of K2-233 d under two different atmospheric scenarios. For an envelope with solar opacity, our solution yielded a core mass of $M_{core} = 5.55\,M_{\oplus}$ and a core radius of $R_{core} = 1.561\,R_{\oplus}$. This corresponds to an envelope radius of $R_{env} = 0.740\,R_{\oplus}$ and an atmospheric mass fraction ($f_{atm}$) of $0.85$\%. Alternatively, an enhanced metallicity model resulted in $M_{core} = 5.56\,M_{\oplus}$ and $R_{core} = 1.562\,R_{\oplus}$, with $R_{env} = 0.739\,R_{\oplus}$ and $f_{atm} = 0.63$\%.

We also model the structure of K2-233 d by assuming an ice-rock interior comprising 25\% ice and 75\% rock and the two different atmospheric scenarios. In this case, for an envelope with solar opacity, our solution yielded a core mass of $M_{core} = 5.59\,M_{\oplus}$ and a core radius of $R_{core} = 1.952\,R_{\oplus}$. This corresponds to an envelope radius of $R_{env} = 0.349\,R_{\oplus}$ and an atmospheric mass fraction ($f_{atm}$) of 0.24\%. Alternatively, an enhanced metallicity model resulted in $M_{core} =5.59\,M_{\oplus}$ and $R_{core} = 1.952\,R_{\oplus}$, with $R_{env} = 0.349\,R_{\oplus}$ and $f_{atm} = 0.18$\%.

Both solutions remain consistent with the planet's observed mass and optical radius. To track the thermal and mass-loss history, we simulated the system’s evolution via photoevaporation. At each time step, we calculated the mass-loss rate to update both $f_{atm}$ and the total planetary mass. The evolving envelope radius was determined using the \cite{Lopez_2014} relation, which, when combined with a constant core radius, provides the time-dependent planetary radius.

Looking toward the future, for Earth-like interior, the planet retains a tenuous envelope with an atmospheric mass fraction between $0.09\%$ and $0.25\%$ and a total radius between $1.713\,R_{\oplus}$ and $1.837\,R_{\oplus}$ (\Autoref{fig:xuv_radius_fenv}). 

For the ice-rock interior, K2-233 d loses its entire envelope within 5 Gyr: at $0.85$ Gyr for solar opacity and $0.71$ Gyr for enhanced opacity. This earlier depletion is primarily a consequence of the lower atmospheric mass fractions at the present ($f_{\text{atm}} = 0.09-0.25\%$) required by the ice-rock structural solution. Since the larger core radius accounts for a greater proportion of the observed total radius, the permitted initial envelope mass is substantially reduced, leading to its rapid exhaustion under photoevaporative flux.

To explore the system's shared history, we simulated the inner planets, K2-233 b and c, assuming they started at 10 Myr with 1\% envelope mass fraction. 
In \Autoref{subsec:rv_analysis} we found very different possible values for planet b's mass.
Therefore, for this planet we simulated two mass core scenarios – $1.9$ $M_{\oplus}$ and $5.2$ $M_{\oplus}$ – corresponding to the smallest and the largest value that we obtained.
Because planet b is situated so close to the host star, in all these two scenarios our models indicate it would be rapidly stripped of this initial atmosphere in just 15 – 70 Myr, leaving behind a bare core consistent with a roughly equal mix of rock and iron. Planet c still completely loses its $1\%$ envelope within 210 to 306 Myr. Therefore, if all three planets formed with $1\%$ initial gas fraction, intense photoevaporation would completely strip b and c well before the system's current age, while allowing the more distant planet d to retain its envelope. This theoretical modelling perfectly aligns with current observational evidence.

\section{Internal composition}
\label{sec:discussion}

\begin{figure}
\centering
    \includegraphics[width=0.9\columnwidth]{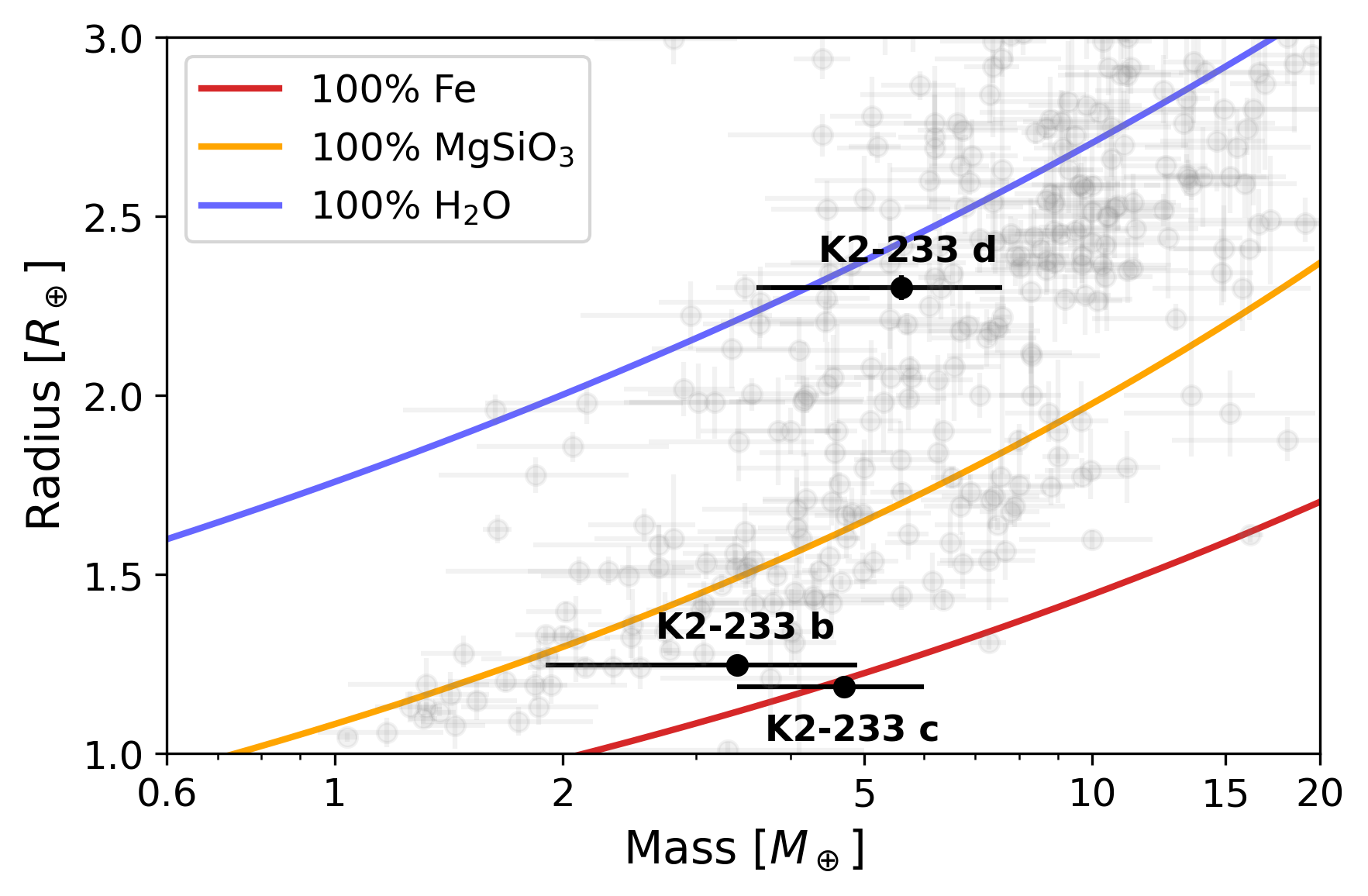}
    \caption{Mass-radius diagram. Black dots show the planets of K2-233, while grey dots are all the other confirmed exoplanets with mass significance greater than $3\sigma$ whose list was downloaded from Nasa Exoplanet Archive \citep{Christiansen_2025}. Coloured curves represent bulk compositions from \cite{Zeng_2019}: pure iron in red, pure silicate in orange, and pure water in blue.}
    \label{sec:data:mass-radius-diagram}
\end{figure}

Based on derived masses and radii of all the planets of K2-233, their bulk densities are $\rho_b = 1.74 \pm 0.78$ $\rho_\oplus$, $\rho_c = 2.82 \pm 0.80$ $\rho_\oplus$, and $\rho_d = 0.46 \pm 0.17$ $\rho_\oplus$ for planets b, c, and d, respectively.
According to the mass-radius diagram in \Autoref{sec:data:mass-radius-diagram}, planets b and c are consistent with predominantly iron compositions. Their positions on the left-hand peak of the radius valley, typically associated with super-Earths, further support this interpretation. 
Planet d, by contrast, lies near the water-world curve in the mass–radius diagram, but its location on the sub-Neptune peak of the radius valley distribution indicates the presence of a significant gaseous envelope.

To further investigate the internal compositions, we developed a custom Bayesian code (Fardella et al. in prep.) based on the framework of \cite{Dorn_2017}, modelling each planet as a four-layer structure: a pure iron core \citep{Hakim_2018}, a silicate mantle \citep{SOTIN2007337}, a water layer \citep{Haldemann_2020}, and a primordial H/He atmosphere \citep{Lopez_2014}. The code determines the masses and radial extent of each component by requiring that their combined mass and radius reproduce the observed within their uncertainties. The code infers the masses of all layers separately.
For each planet we considered both the smallest and the largest mass values physically plausible that we obtained in \Autoref{subsec:rv_analysis} and performed two different simulations to model the interior, one for each mass value. 
Posterior distributions of the composition models for planets b, c, and d are presented in \Autoref{appendix:figure_tables}.

For planet b the smallest and largest masses are $1.9^{+1.3}_{-1.1}\ M_\oplus$ and $5.2 \pm 1.5\ M_\oplus$, respectively.
In both the scenarios the solutions suggest that the interior is dominated by the core ($M_\text{core} = 1.35^{+0.25}_{-0.33}$ $M_\oplus$ and $M_\text{core} = 4.25^{+0.45}_{-0.50}$ $M_\oplus$, in the case with total mass $1.9^{+1.3}_{-1.1}$ $M_\oplus$ and $5.2 \pm 1.5$ $M_\oplus$, respectively), with possible minor presence of the mantle and water layers which in both cases have been retrieved as upper limits respectively of $M_\text{man} < 0.77$ $M_\oplus$ and $M_\text{water} < 0.32$ $M_\oplus$ at $1\sigma$ in the case with total mass equal to $1.9^{+1.3}_{-1.1}$ $M_\oplus$, and of $M_\text{man} < 1.24$ $M_\oplus$ and $M_\text{water} < 0.32$ $M_\oplus$ for the case with total mass equal to $5.2 \pm 1.5$ $M_\oplus$.
In both cases the gas envelope is negligible, with a mass of $M_\text{gas} \sim 2 \times 10^{-5}$ $M_\oplus$.

For planet c, the two considered total masses are $4.2^{+1.6}_{-1.4}\ M_\oplus$ and $4.7 \pm 1.3\ M_\oplus$. In both cases, the inferred structure is core-dominated, with core masses of $M_\text{core} = 3.91^{+0.18}_{-0.35}\ M_\oplus$ and $4.46^{+0.15}_{-0.28}\ M_\oplus$, respectively. For the lower-mass configuration, the mantle and water layers are constrained to $M_\text{man} < 0.58\ M_\oplus$ and $M_\text{water} < 0.12\ M_\oplus$ at $1\sigma$, respectively, while the corresponding upper limits for the higher-mass configuration are $M_\text{man} < 0.47\ M_\oplus$ and $M_\text{water} < 0.09\ M_\oplus$. In both cases, the H/He envelope is negligible, with a mass of $M_\text{gas} \sim 1 \times 10^{-5}\ M_\oplus$.

For planet d, the lowest and highest masses considered are $2.6^{+2.5}_{-1.6}\ M_\oplus$ and $7.8 \pm 1.9\ M_\oplus$, respectively. In the lower-mass configuration, K2-233 d is consistent with a water-rich composition, with a well-constrained water mass of $M_\text{water} = 1.16 \pm 0.08\ M_\oplus$ and a negligible H/He envelope ($M_\text{gas} \sim 3 \times 10^{-7}\ M_\oplus$). The corresponding core and mantle masses are $M_\text{core} = 0.99^{+0.21}_{-0.22}\ M_\oplus$ and $M_\text{man} = 0.43^{+0.27}_{-0.22}\ M_\oplus$, respectively.
The structure inferred for the higher-mass configuration ($7.8 \pm 1.9\ M_\oplus$) is markedly different. The core and mantle masses are $M_\text{core} = 4.86^{+1.13}_{-1.01}\ M_\oplus$ and $M_\text{man} = 1.00^{+0.96}_{-0.62}\ M_\oplus$, respectively, while the water mass is unconstrained owing to the degeneracy between the different layers. In contrast, the H/He envelope is well constrained, with $\log M_{\text{gas}} = -1.69^{+0.22}_{-0.31}$, corresponding to $M_\text{gas} \sim 0.02\ M_\oplus$ and an atmospheric thickness of $R_{\text{gas}} \sim 0.41\ R_\oplus$. This configuration therefore favours a substantial primordial H/He envelope.
The two mass assumptions thus lead to substantially different interior structures, ranging from a water-rich planet with a negligible atmosphere to a sub-Neptune with a significant H/He envelope. However, according to the MGIC (see \Autoref{tab:several-analyses}), the Multi-GP framework yielding $M_d = 2.6^{+2.5}_{-1.6}\ M_\oplus$ is the least favoured. We therefore regard the corresponding water-rich solution as less likely, and favour an interpretation in which K2-233 d retains a substantial primordial H/He envelope.
We also modelled the interior of K2-233 d using the mass derived from Model 9. The resulting structure is very similar to that obtained for $M_d = 7.8 \pm 1.9\ M_\oplus$: the water mass remains unconstrained, while a substantial primordial atmosphere is inferred, with $\log M_{\text{gas}} = -1.70^{+0.22}_{-0.34}$ ($M_\text{gas} \sim 0.02\ M_\oplus$).
Although interior structure models can present degeneracies between water-rich compositions and H/He envelopes, the position of K2-233 d on the sub-Neptune peak of the radius valley distribution and the results of the photoevaporation modelling both favour the presence of a primordial H/He atmosphere.
This atmospheric thickness differs from that derived in \Autoref{subsec:planet-d-photoevaporation}. The discrepancy arises from the different modelling approaches adopted for the solid interior. In the present section, the core and mantle structure are computed by directly solving the equations of state, providing a detailed description of the internal layers at the expense of a higher computational cost. In \Autoref{subsec:planet-d-photoevaporation}, by contrast, the planet had to be evolved over its entire history, requiring thousands of simulations. A full equation-of-state treatment would therefore have been computationally prohibitive. For that purpose, we adopted the analytical relations of \cite{Fortney2007} to estimate the core properties.
In both approaches, the total planetary mass and radius are constrained to reproduce the observed values. Consequently, differences in the inferred core parameters propagate into differences in the atmospheric properties. Moreover, the approach used in \Autoref{subsec:planet-d-photoevaporation} does not include a water layer, which in the model adopted in this section contributes to the total mass and radius of the planet. Although this component is not strongly constrained for K2-233 d, both approaches consistently indicate the presence of a substantial primordial atmosphere. Therefore, the assumptions adopted in \Autoref{subsec:planet-d-photoevaporation} appear reasonable, and the results of that analysis are reaffirmed.

\section{Summary and conclusions}
\label{sec:conclusions}

We presented the re-analysis of the K2-233 system, taking the advantage of additional HARPS-N spectra obtained within the ArMS programme, combined with the HARPS and K2 archival data, the TESS light curves collected during a portion of the HARPS-N monitoring.
Due to the high activity level of the host star, we explored several Multi-GP configurations and demonstrated that the inferred planetary masses are strongly dependent on the chosen activity-modelling strategy, yielding a set of different physically viable solutions rather than a single uniquely preferred model.

Planet d lies on the right-hand peak of the radius valley distribution at $\sim 2.4$ $R_\oplus$ and has a bulk density of $\rho_d = 0.46 \pm 0.17$ $\rho_\oplus$, indicating the presence of a substantial primordial atmosphere. We investigated the impact of the host star’s XUV irradiation on planet d's atmospheric evolution by modelling the system from 10 Myr to 5 Gyr under photoevaporative mass loss.

Our photoevaporation models indicate that the long-term evolution of K2-233 d is strongly dependent on its internal structure. For an Earth-like interior, the planet retains a tenuous atmosphere at 5 Gyr, with an atmospheric mass fraction of $0.09-0.25\%$ and a final radius of $1.713-1.837\ R_\oplus$. Conversely, for the ice-rock interior, the envelope is completely stripped within $0.71 - 0.85$ Gyr, leaving a bare core with a radius of $1.952\ R_\oplus$. K2-233 d therefore represents a transitional object whose long-term fate depends on the assumed interior composition and atmospheric properties, either retaining a low-mass envelope within the radius valley or evolving into a bare core.

Our models indicate that, if K2-233 b and c initially possessed primordial gaseous envelopes similar to that assumed for K2-233 d at 10 Myr, intense photoevaporation would have removed them on relatively short timescales. In particular, the envelopes of planets b and c would be completely stripped within $15$ – $70$ Myr and in the range $210$ – $306$ Myr, respectively, leaving behind predominantly iron and rocky cores, consistent with the compositions inferred from our interior-structure modelling. These results support a common evolutionary scenario in which the different present-day properties of the three planets are primarily driven by atmospheric escape.

We modelled the interiors of the planets finding that planets b and c are iron-dominated with possible minor presence of silicate and water, while planet d presents a substantial H/He atmosphere with the possible presence of water.

As a final remark, we note that K2-233 d emerges as a highly compelling candidate for atmospheric characterisation with the JWST, drawing a natural parallel to the benchmark sub-Neptune K2-18 b \citep{Madhusudhan_2025_k218}. 
Assuming a hydrogen-dominated atmosphere, we estimate an atmospheric scale height of $H = 155 \pm 49$ km, indicative of an extended atmosphere. The corresponding Transmission Spectroscopy Metric (TSM; \citealt{Kempton_2018}) is $\mathrm{TSM} = 41.2 \pm 13.2$.
Although K2-233 d presents a smaller planet-to-star radius ratio ($R_p/R_{\text{star}} = 0.03048 \pm 0.00043$) – yielding a shallower baseline transit depth – this geometric limitation is effectively offset by the superior infrared magnitude of its host star, which guarantees the high photon flux necessary to achieve excellent SNR. Furthermore, the planet's slightly higher equilibrium temperature ($T_\text{eq} = 731 \pm 9$ K) provides a distinct observational advantage: it mitigates the risk of widespread condensation and pervasive cloud deck formation, ensuring that key volatile species remain in the gas phase and accessible to transmission spectroscopy.

\begin{acknowledgements}

The authors acknowledge the anonymous Referee for her/his
useful comments and suggestions.

The authors acknowledge the support from ASI-INAF agreement 2021-5-HH.0 and the INAF Guest Observer Grant (Normal) "ArMS: the Ariel Masses Survey Large Program at the TNG" (Prog ID: \texttt{AOT48TAC\_48}). RS acknowledges the support from the European Union – Next Generation EU through the grant n. 2022J7ZFRA – Exo-planetary Cloudy Atmospheres and Stellar High energy (Exo-CASH) funded by MUR – PRIN 2022. 

We acknowledge support from the European Union – NextGenerationEU (PRIN MUR 2022 20229R43BH) and the “Programma di Ricerca Fondamentale INAF 2023”. We acknowledge financial contribution from the INAF Large Grant 2023 “EXODEMO”.

We acknowledge the Italian center for Astronomical Archives (IA2, \url{https://www.ia2.inaf.it}), part of the Italian National Institute for Astrophysics (INAF), for providing technical assistance, services and supporting activities of the GAPS collaboration.

This work has made use of data from the European Space Agency (ESA) mission {\it Gaia} (\url{https://www.cosmos.esa.int/gaia}), processed by the {\it Gaia} Data Processing and Analysis Consortium (DPAC, \url{https://www.cosmos.esa.int/web/gaia/dpac/consortium}). Funding for the DPAC has been provided by national institutions, in particular the institutions participating in the {\it Gaia} Multilateral Agreement.

This paper includes data collected by the TESS mission, which are publicly available from the Mikulski Archive for Space Telescopes (MAST). Funding for the TESS mission is provided by NASA's Science Mission directorate.

This research made use of Lightkurve, a Python package for Kepler and TESS data analysis \citep{lightkurve_2018}.

The research activities described in this paper were carried out with contribution of the Next Generation EU funds within the National Recovery and Resilience Plan (PNRR), Mission 4 - Education and Research, Component 2 - From Research to Business (M4C2), Investment Line 3.1 - Strengthening and creation of Research Infrastructures, Project IR0000034 – “STILES - Strengthening the Italian Leadership in ELT and SKA.

\end{acknowledgements}

\bibliographystyle{aa}
\bibliography{bibliography}

\begin{appendix}

\section{Detection Completeness}
\label{appendix:detection_compl}

\begin{figure}
\centering
    \includegraphics[width=\columnwidth]{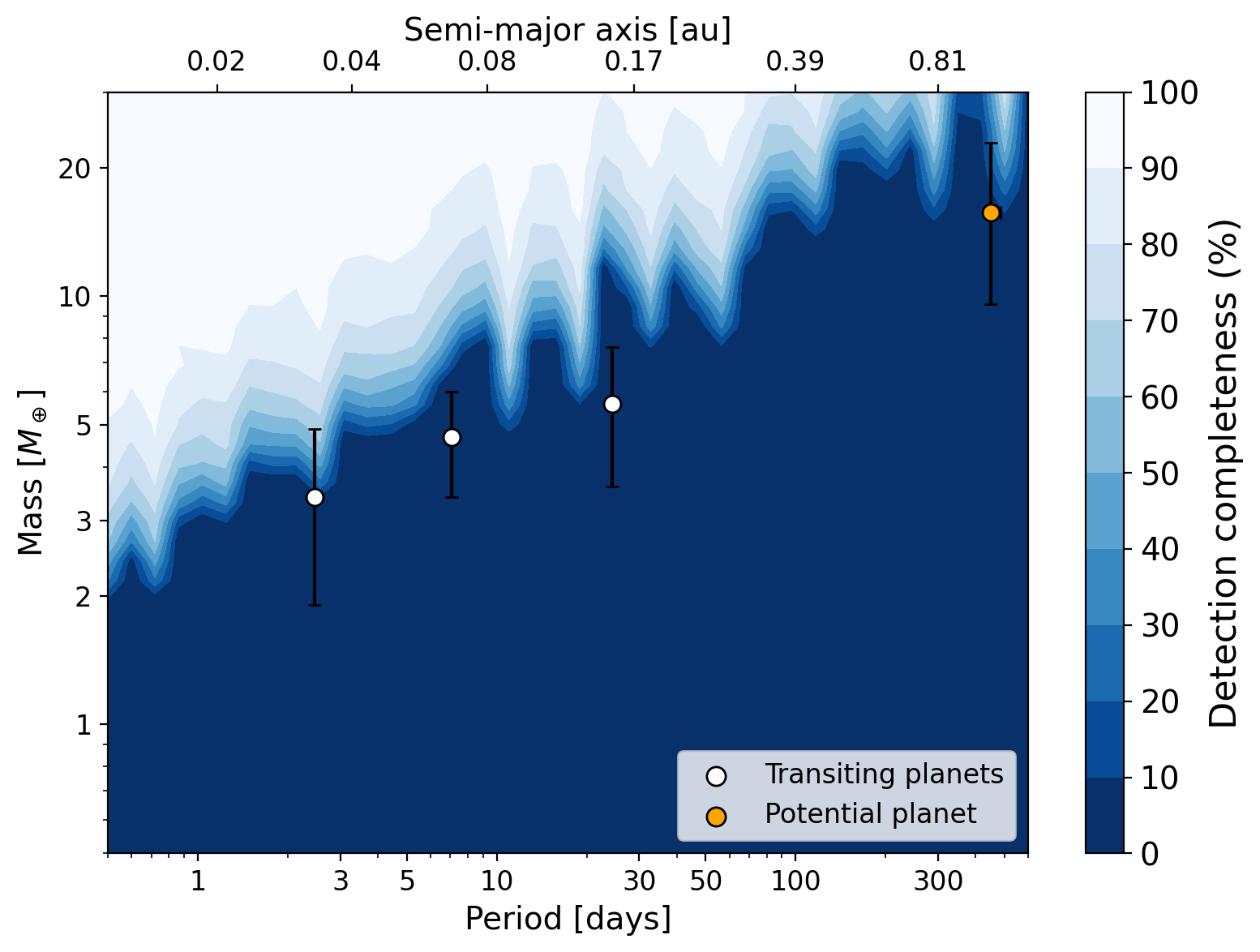}
    \caption{Detection completeness map of the K2-233 RV time series. The colour scale indicates the recovery rate (from $0\%$ to $100\%$) of synthetic signals injected into the RV residuals of the three-planets model. White circles represent the three known transiting planets with the masses obtained from \Autoref{subsec:rv_analysis}, while the orange circle indicates the candidate signal at $P \sim 449$ days.}
    \label{sec:data:detection-completeness}
\end{figure}

To assess the RV time series detection completeness we perform an injection and retrieval scheme in which synthetic companions are injected into the post-fit residuals of the three-planets configuration and they are subjected to statistical tests to assess their detectability. For this purpose we follow an approach used in previous works \citep[e.g.][]{barbato2018, barbato2025, gratton2024, matthews2023, Ruggieri2026}. We explore a grid of $40 \times 40$ companion minimum masses and orbital periods ranging from $0.5$ $M_\oplus$ to $0.1$ $M_J$ and from $0.1$ to $600$ days. For each mass-period realisation we generate 1000 synthetic RV curves corresponding to the injected companion with randomly drawn values of $e$, $\omega$, $\lambda$, considering each of the $1.6 \times 10^6$ injected signals as detected if the corresponding FAP value is $\leq10^{-3}$. We finally compute the detection completeness of each mass-period realisation as the ratio between detectable injected signals and total injections, obtaining the detection completeness map shown in \Autoref{sec:data:detection-completeness}.

\section{Additional figures and tables}
\label{appendix:figure_tables}

\begin{figure}
\centering
    \includegraphics[width=\columnwidth]{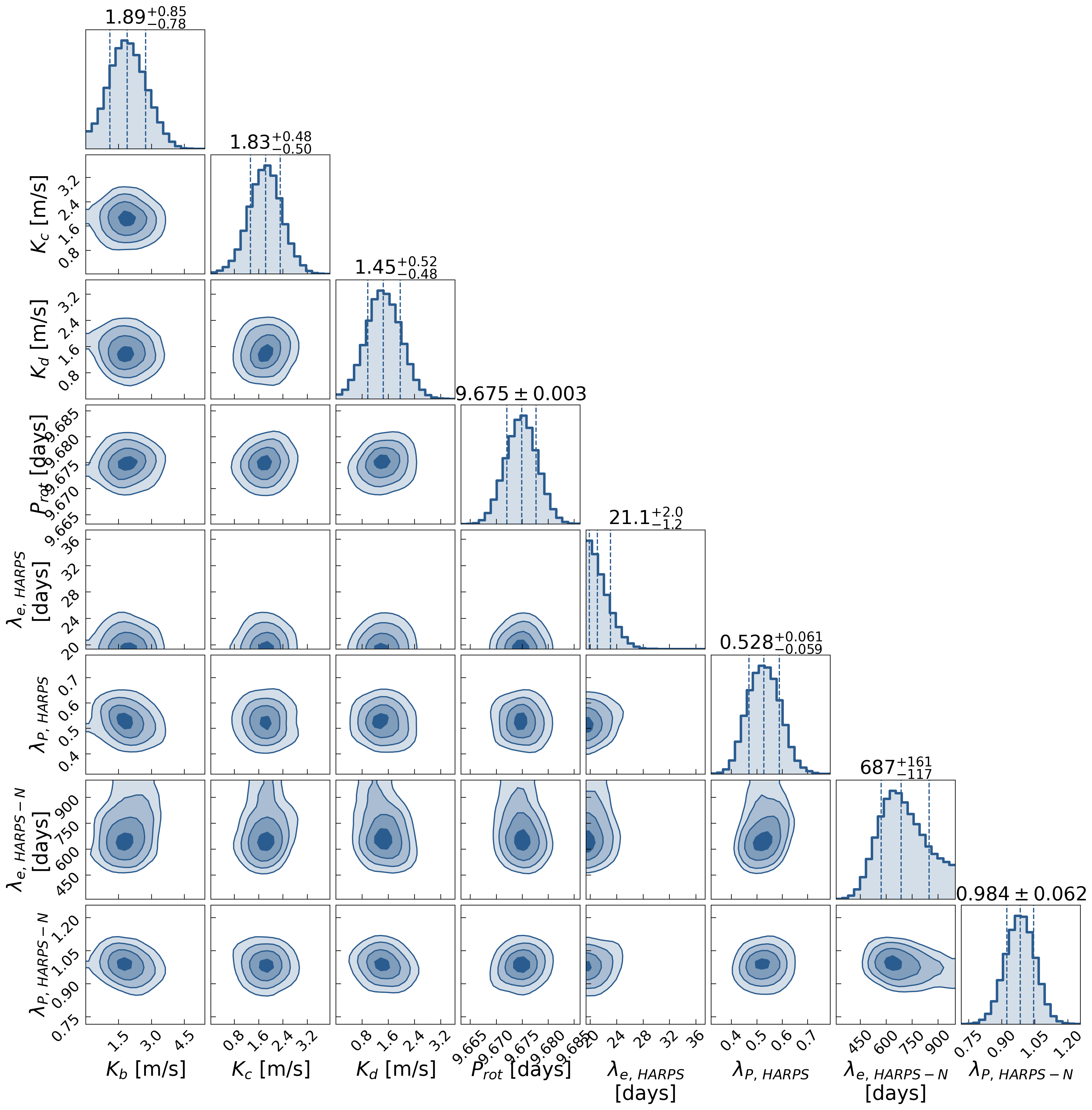}
    \caption{Posterior distribution functions of the planets RV semi-amplitudes and the hyperparameters of the RV's GP retrieved from the nested sampler for the models analysed in \Autoref{subsec:rv_analysis} for the circular-orbit configuration. The 2D plots represent the correlation between the parameters.}
    \label{sec:data:corners}
\end{figure}

\begin{figure}
\centering
    \includegraphics[width=.7\columnwidth]{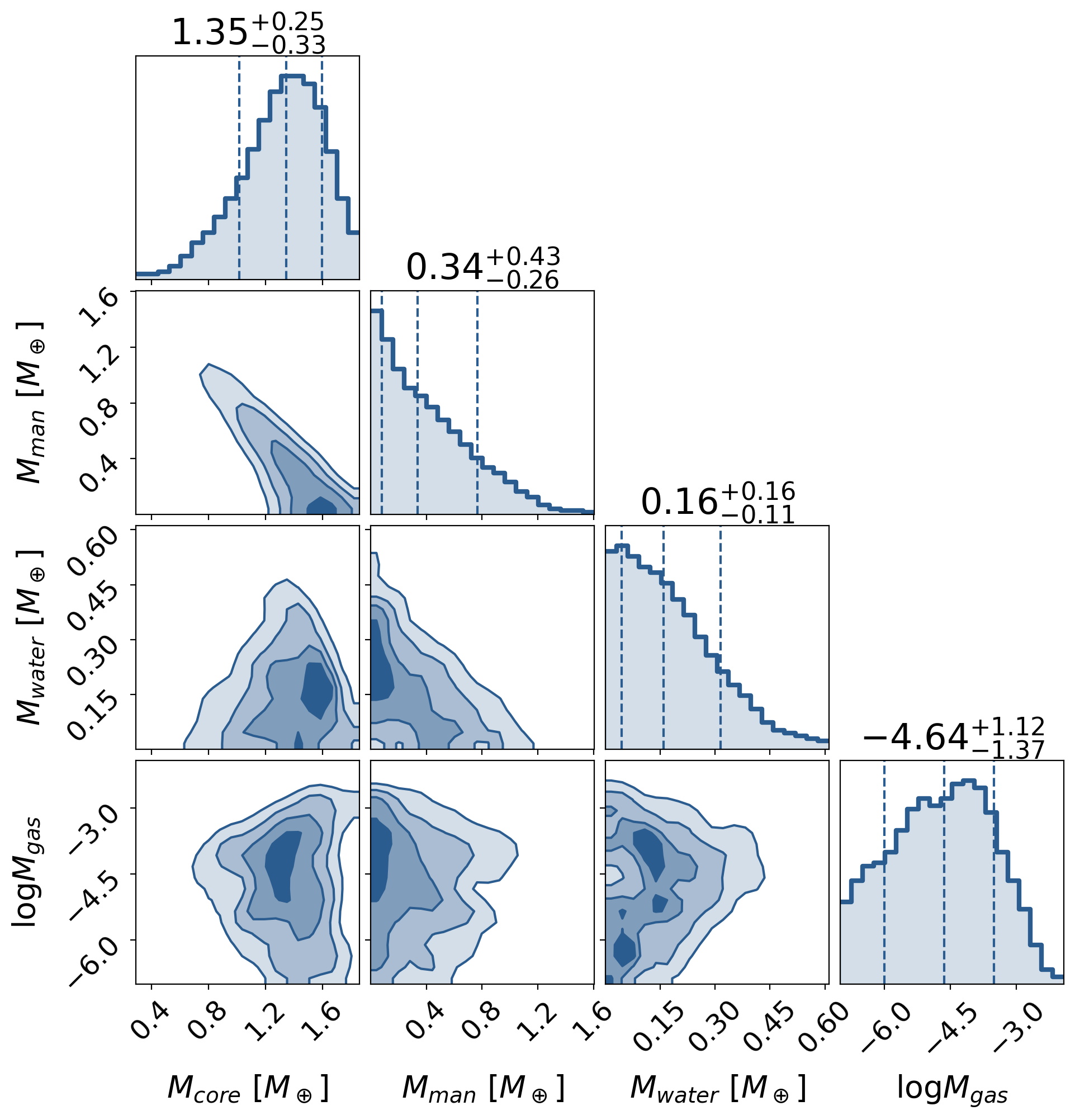}
    \caption{Posterior probability distributions of the masses inferred to model the interior structure of planet b assuming total mass equal to $M_b = 1.9^{+1.3}_{-1.1}$ $M_\oplus$. From left to right, the panels show the core, the mantle, and the water masses, and the logarithm of the H/He envelope mass.}
    \label{sec:data:PANELLA-b-small}
\end{figure}

\begin{figure}
\centering
    \includegraphics[width=.7\columnwidth]{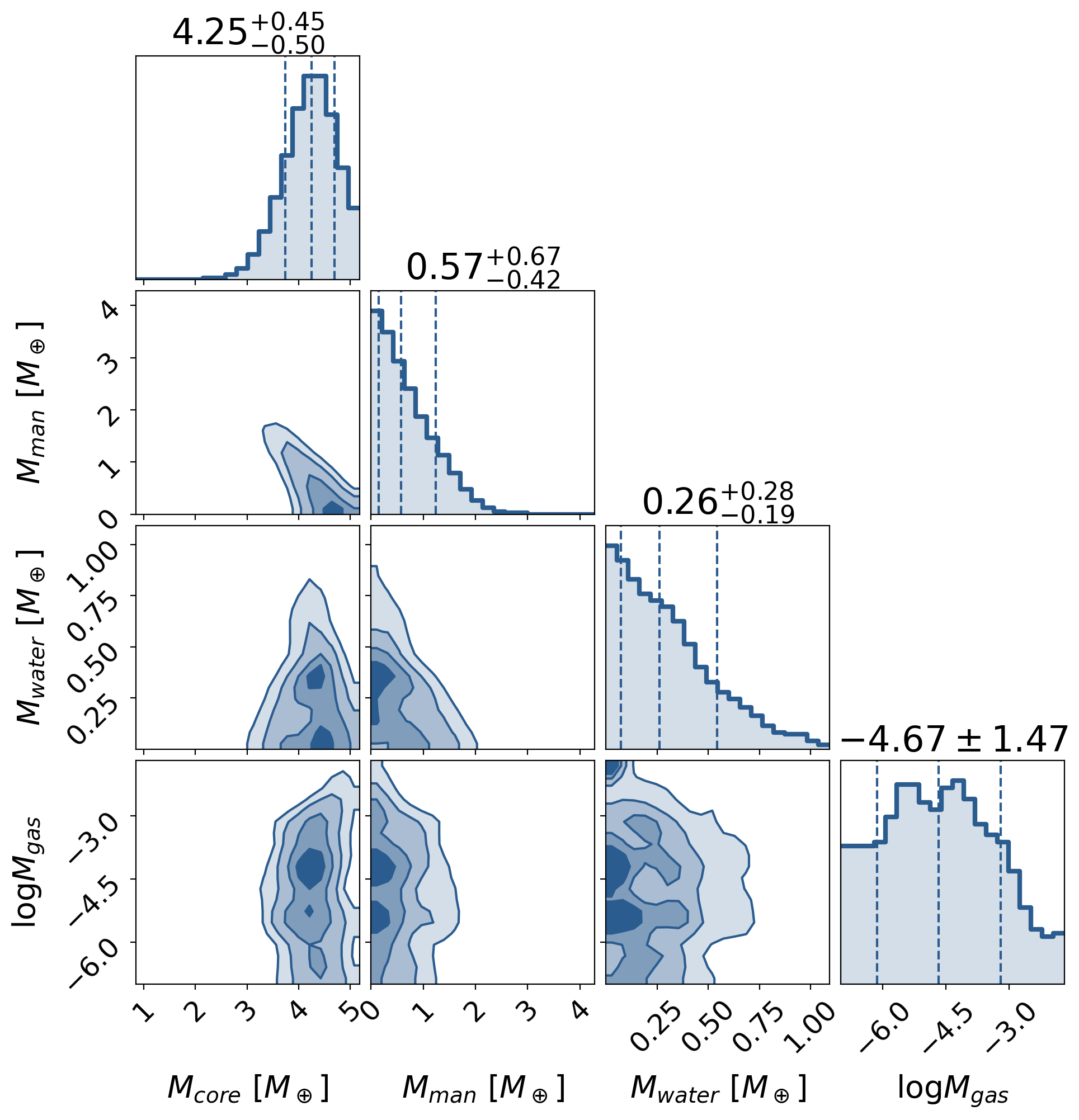}
    \caption{Same of \Autoref{sec:data:PANELLA-b-small}, but with total mass equal to $M_b = 5.2 \pm 1.5$ $M_\oplus$.}
    \label{sec:data:PANELLA-b-large}
\end{figure}

\begin{figure}
\centering
    \includegraphics[width=.7\columnwidth]{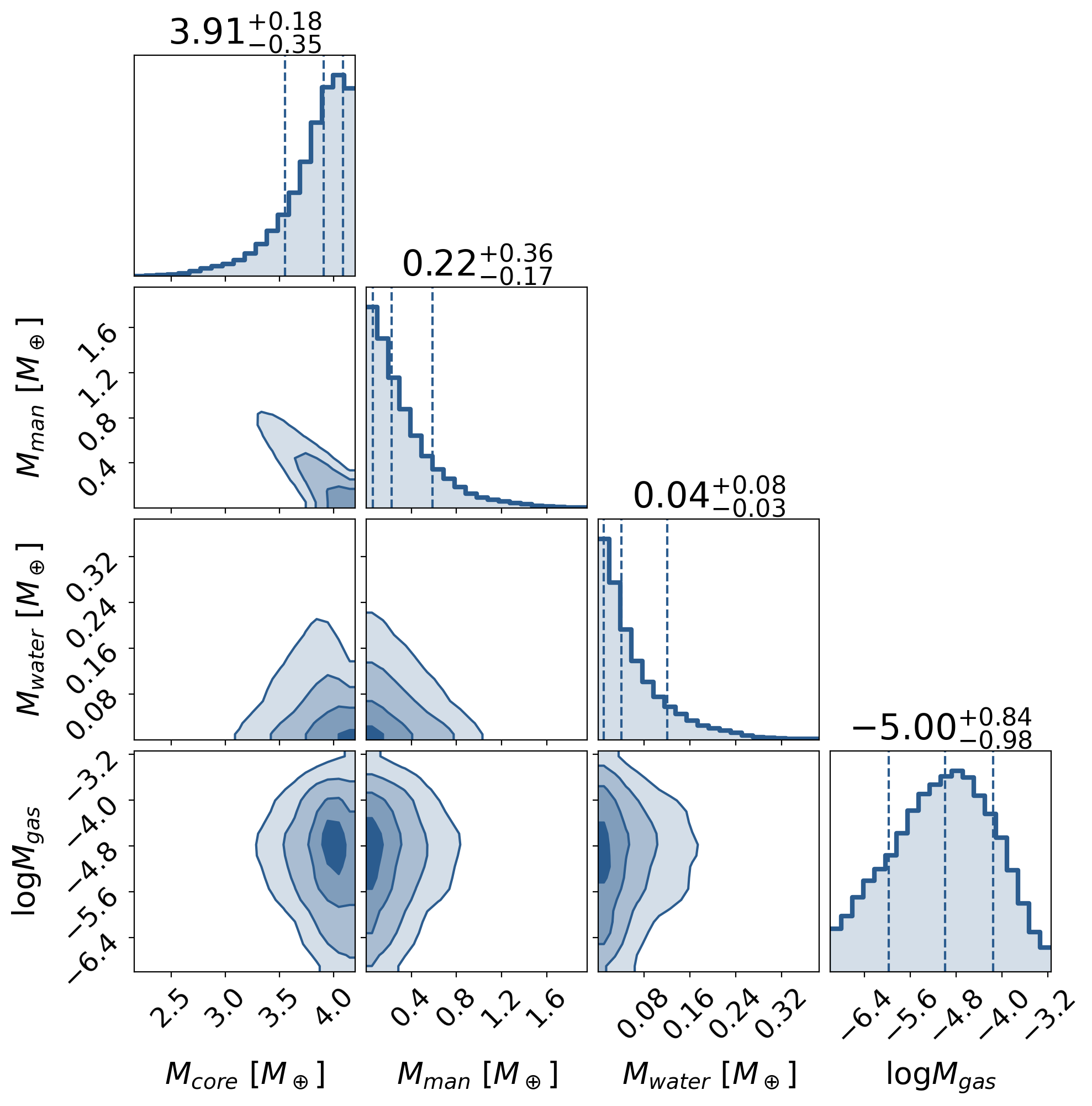}
    \caption{Same of \Autoref{sec:data:PANELLA-b-small}, but for planet c with total mass equal to $M_c = 4.2^{+1.6}_{-1.4}\ M_\oplus$.}
    \label{sec:data:PANELLA-c-small}
\end{figure}

\begin{figure}
\centering
    \includegraphics[width=.7\columnwidth]{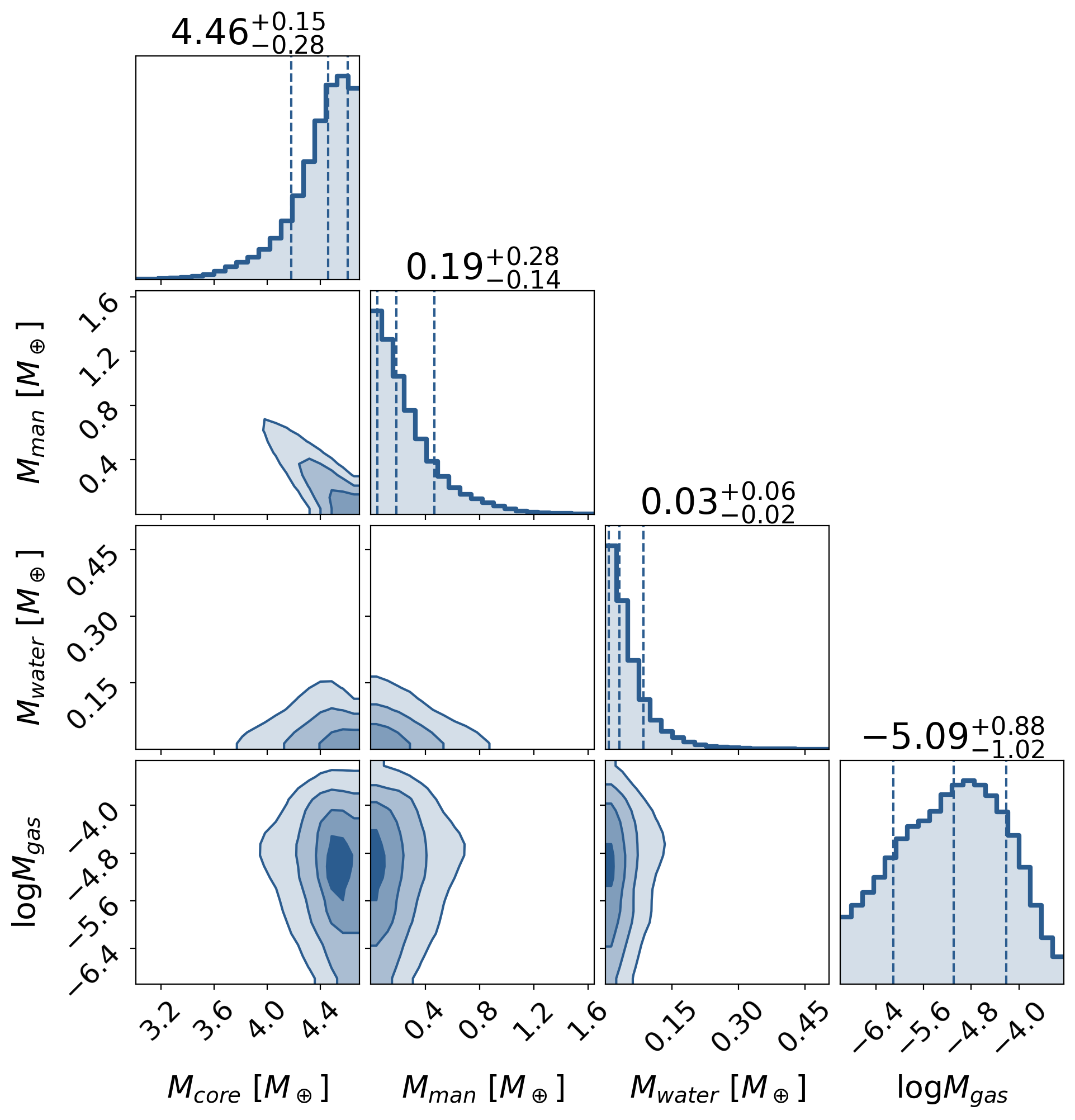}
    \caption{Same of \Autoref{sec:data:PANELLA-b-small}, but for planet c with total mass equal to $M_c = 4.7 \pm 1.3\ M_\oplus$.}
    \label{sec:data:PANELLA-c-large}
\end{figure}

\begin{figure}
\centering
    \includegraphics[width=.7\columnwidth]{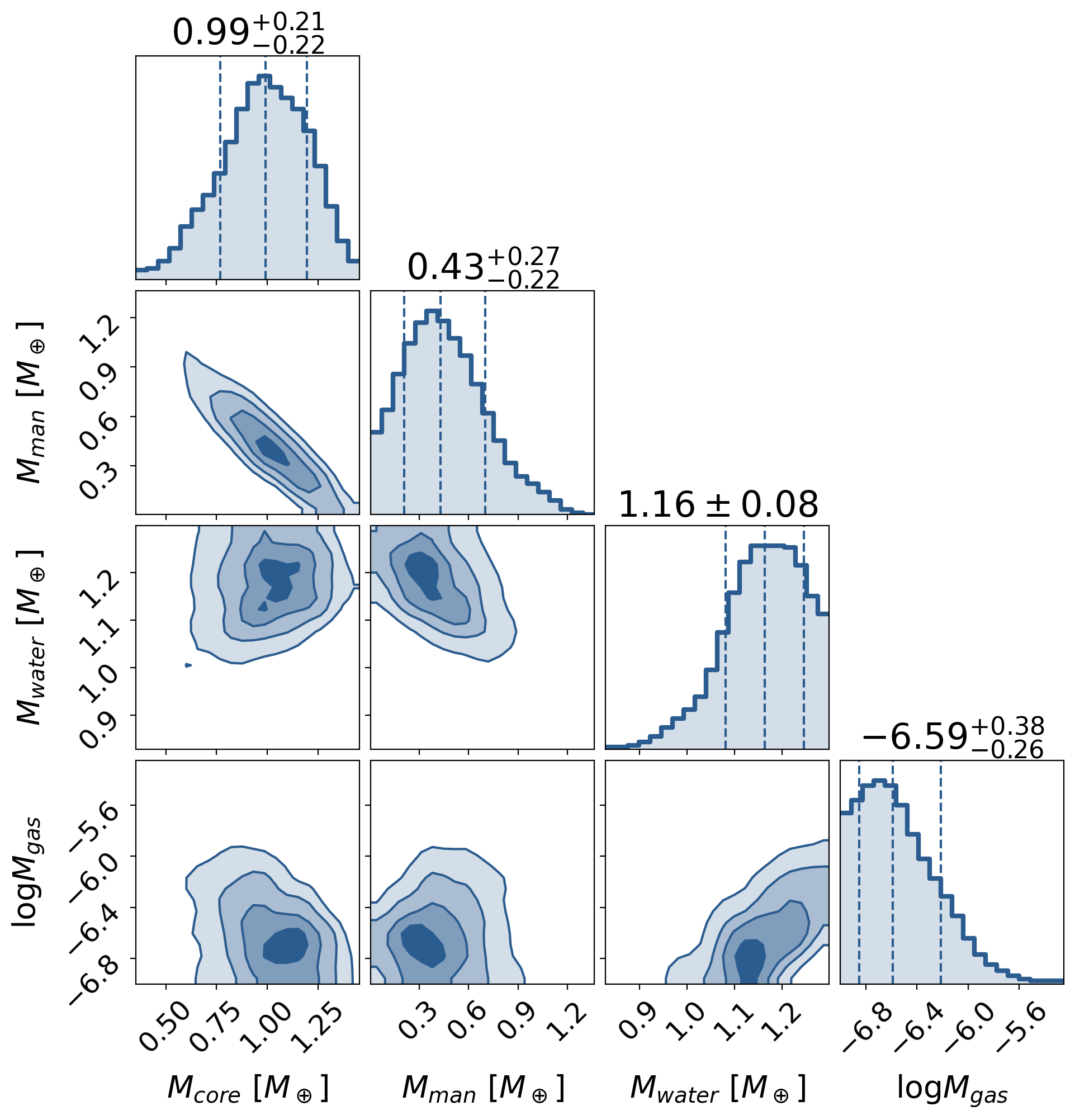}
    \caption{Same of \Autoref{sec:data:PANELLA-b-small}, but for planet d with total mass equal to $M_d = 2.6^{+2.5}_{-1.6}\ M_\oplus$.}
    \label{sec:data:PANELLA-d-small}
\end{figure}

\begin{figure}
\centering
    \includegraphics[width=.7\columnwidth]{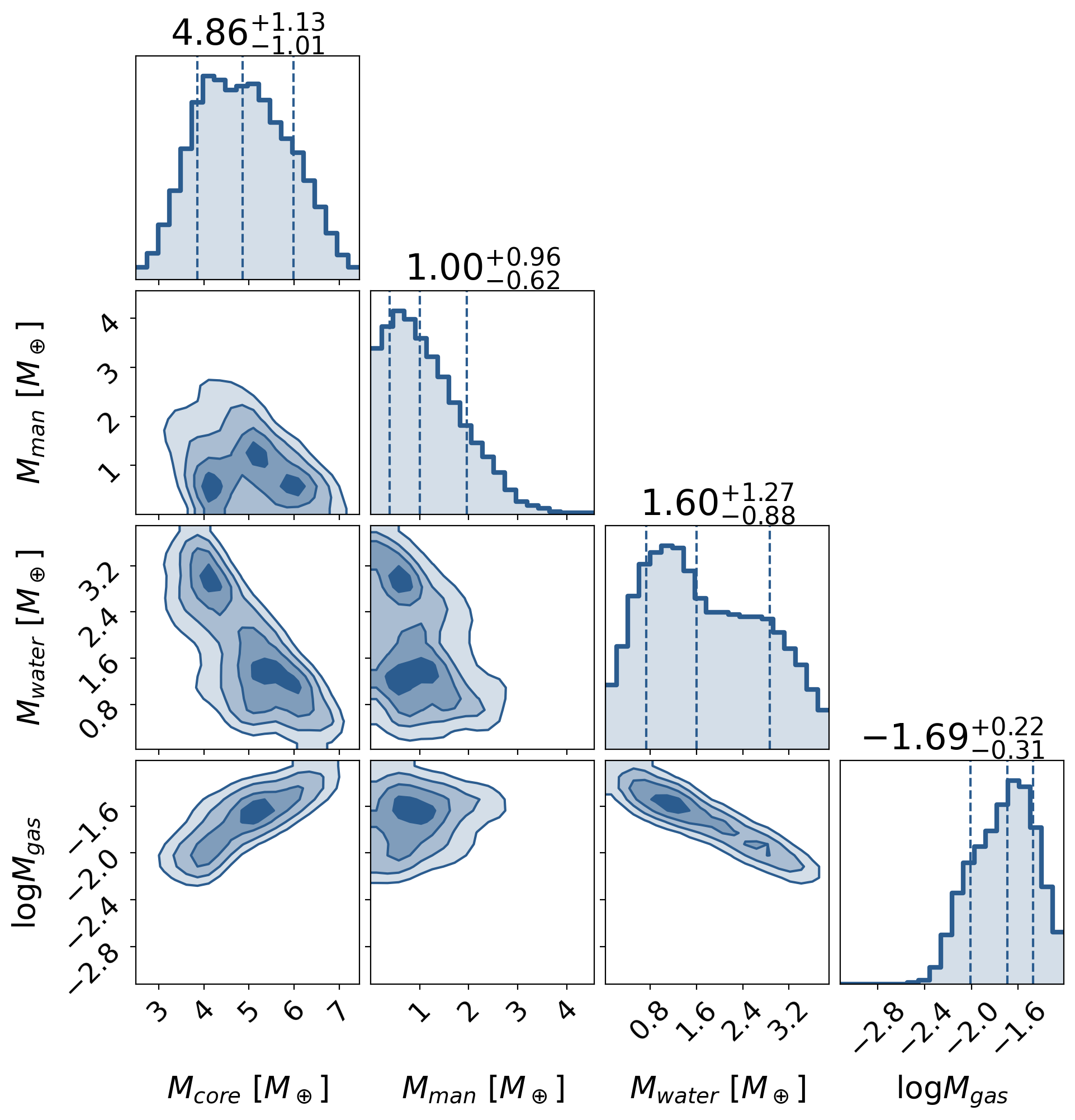}
    \caption{Same of \Autoref{sec:data:PANELLA-b-small}, but for planet d with total mass equal to $M_d = 7.8 \pm 1.9\ M_\oplus$.}
    \label{sec:data:PANELLA-d-large}
\end{figure}

\begin{figure*}
\centering
    \includegraphics[width=\textwidth]{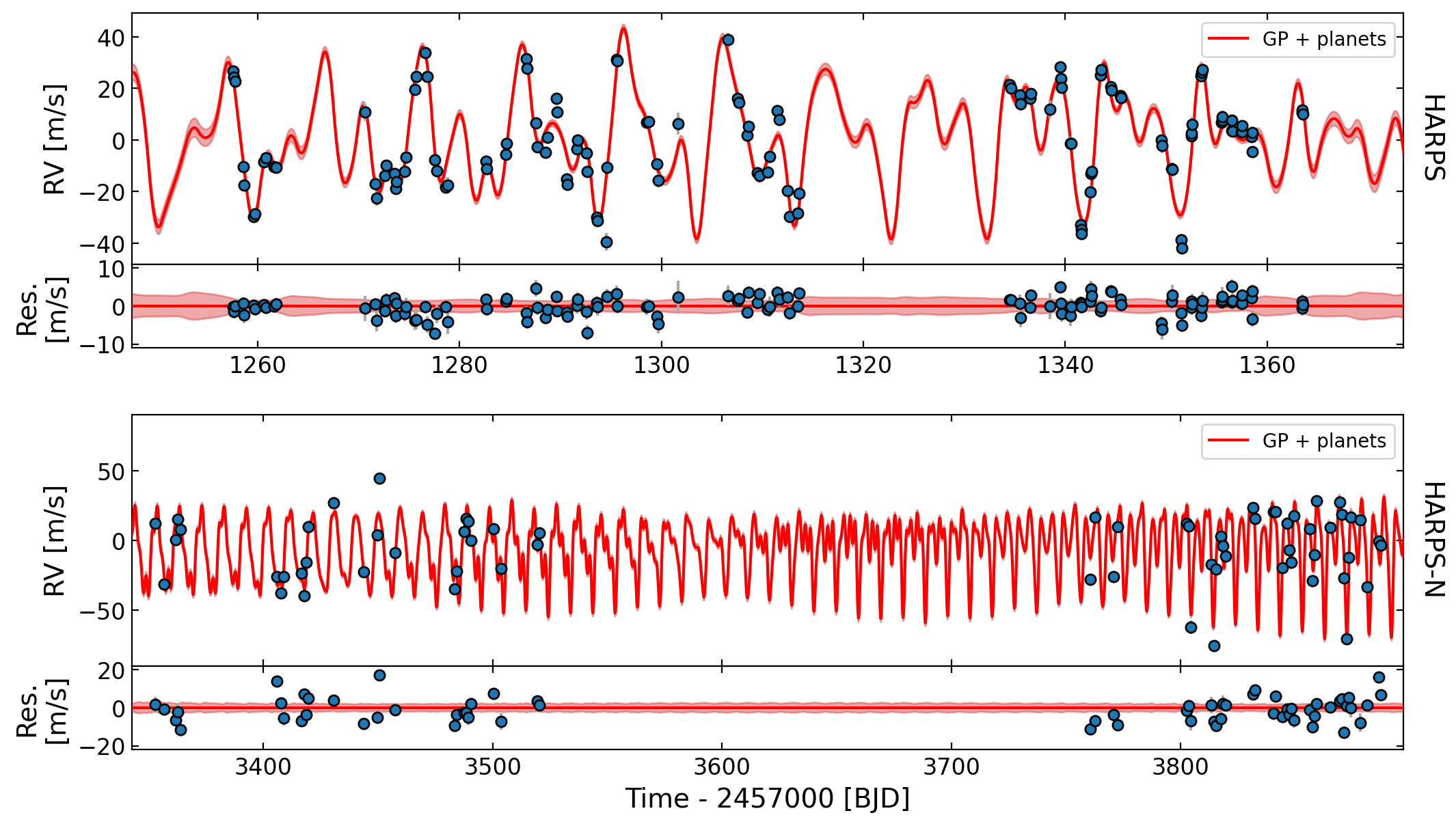}
    \caption{RV time series and best-fitting GP and planetary models from Model 9 in the circular-orbit configuration. The upper panels show the HARPS RV measurements and the corresponding residuals, while the lower panels show the HARPS-N RV measurements and residuals. The solid red lines represent the total model, comprising the GP activity component and the three planetary Keplerian signals.}
    \label{sec:data:rv-gp-model}
\end{figure*}

In this appendix, \Autoref{sec:data:corners} shows the posterior distributions of the planetary RV semi-amplitudes inferred from the Model 9 described in \Autoref{subsec:rv_analysis} along with the respective GP's hyperparameters of the RV dataset.
In \Autoref{sec:data:rv-gp-model} we present the RV time series together with the GP model and the Keplerian signals of the planets.
\Autoref{sec:data:PANELLA-b-small}, \Autoref{sec:data:PANELLA-b-large}, \Autoref{sec:data:PANELLA-c-small}, \Autoref{sec:data:PANELLA-c-large}, \Autoref{sec:data:PANELLA-d-small}, and \Autoref{sec:data:PANELLA-d-large} show the posterior distributions of parameters implemented to model the inner composition of planet b, c, and d.
\Autoref{sec:table:priors} lists the prior distributions adopted for all model parameters together with their derived values.

\onecolumn

\begingroup

\setlength{\LTcapwidth}{\textwidth}

\setlength{\LTleft}{0pt}
\setlength{\LTright}{0pt}

\begin{longtable}{@{\extracolsep{\fill}}lcccc}
  \caption[]{Priors and posteriors distributions of models' parameters used for the analyses of K2-233.}
  \label{sec:table:priors} \\

  \toprule
  \toprule
  & \multicolumn{2}{c}{\textbf{Light curve}} & \multicolumn{2}{c}{\textbf{Radial velocity (Model 9)}} \\
  \midrule
  Parameter & Prior & Final value & Prior & Final value \\
  \midrule
  \endfirsthead

  \caption[]{(Continued)} \\
  \toprule
  \toprule
  & \multicolumn{2}{c}{\textbf{Light curve}} & \multicolumn{2}{c}{\textbf{Radial velocity (Model 9)}} \\
  \midrule
  Parameter & Prior & Final value & Prior & Final value \\
  \midrule
  \endhead

  \bottomrule
  \multicolumn{5}{r@{}}{Continued on next page} \\
  \endfoot

  \bottomrule
  \endlastfoot

  \textbf{Planet b's parameters} & \multicolumn{4}{c}{} \\[3pt]
  $P_b$ [days] & 
  $\mathcal{N}[2.46750,\,0.0001]$ & $2.4674559^{+0.0000039}_{-0.0000044}$ & 
  $\mathcal{N}[2.4674559,\,0.0000044]$ & $2.4674559^{+0.0000036}_{-0.0000033}$ \\[3pt]
  $T_{0, b}$ [BJD $-$ 2457000] & 
  $\mathcal{N}[991.69112,\,0.001]$ & $991.69124^{+0.00054}_{-0.00043}$ & 
  $\mathcal{N}[991.69124,\,0.00054]$ & $991.69130 \pm 0.00044$ \\[3pt]
  $i_b$ [deg] & 
  $\mathcal{U}[75,\,90]$ & $89.45^{+0.39}_{-0.51}$ & 
  $-$ & $-$ \\[3pt]
  $R_b/R_{\text{star}}$ & 
  $\mathcal{U}[0.00,\,0.05]$ & $0.01653^{+0.00026}_{-0.00025}$ & 
  $-$ & $-$ \\[3pt]
  $e_b$ & 
  $\mathcal{F}[0]$ & $0$ & 
  $\mathcal{F}[0]$ & $0$ \\[3pt]
  $\omega_b$ [deg] & 
  $\mathcal{F}[90]$ & $90$ & 
  $\mathcal{F}[90]$ & $90$ \\[3pt]
  $K_b$ [m s$^{-1}$] & 
  $-$ & $-$ & 
  $\mathcal{U}[0,\,10]$ & $1.89^{+0.85}_{-0.78}$ \\[3pt]
  \multicolumn{5}{c}{} \\[3pt]
  
  \textbf{Planet c's parameters} & \multicolumn{4}{c}{} \\[3pt]
  $P_c$ [days] & 
  $\mathcal{N}[7.06,\,0.01]$ & $7.060308^{+0.000018}_{-0.000034}$ & 
  $\mathcal{N}[7.060308,\,0.000034]$ & $7.05972^{+0.00029}_{-0.00037}$ \\[3pt]
  $T_{0, c}$ [BJD $-$ 2457000] & 
  $\mathcal{N}[586.88,\,0.01]$ & $586.8613^{+0.0023}_{-0.0017}$ & 
  $\mathcal{N}[586.8613,\,0.0023]$ & $586.8631^{+0.0015}_{-0.0013}$ \\[3pt]
  $i_c$ [deg] & 
  $\mathcal{U}[75,\,90]$ & $89.77^{+0.16}_{-0.22}$ & 
  $-$ & $-$ \\[3pt]
  $R_c/R_{\text{star}}$ & 
  $\mathcal{U}[0.00,\,0.05]$ & $0.01571^{+0.00034}_{-0.00033}$ & 
  $-$ & $-$ \\[3pt]
  $e_c$ &
  $\mathcal{F}[0]$ & $0$ & 
  $\mathcal{F}[0]$ & $0$ \\[3pt]
  $\omega_c$ [deg] & 
  $\mathcal{F}[90]$ & $90$ & 
  $\mathcal{F}[90]$ & $90$ \\[3pt]
  $K_c$ [m s$^{-1}$] & 
  $-$ & $-$ & 
  $\mathcal{U}[0,\,10]$ & $1.83^{+0.48}_{-0.50}$ \\[3pt]
  \multicolumn{5}{c}{} \\[3pt]
  
  \textbf{Planet d's parameters} & \multicolumn{4}{c}{} \\[3pt]
  $P_d$ [days] & 
  $\mathcal{N}[24.36,\,0.01]$ & $24.36491^{+0.00055}_{-0.00057}$ & 
  $\mathcal{N}[24.36491,\,0.00057]$ & $24.36455^{+0.00024}_{-0.00035}$ \\[3pt]
  $T_{0, d}$ [BJD $-$ 2457000] & 
  $\mathcal{N}[1005.58,\,0.01]$ & $1005.58277^{+0.00098}_{-0.00090}$ & 
  $\mathcal{N}[1005.58277,\,0.00098]$ & $1005.58332^{+0.00063}_{-0.00052}$ \\[3pt]
  $i_d$ [deg] & 
  $\mathcal{U}[75,\,90]$ & $89.48^{+0.048}_{-0.043}$ & 
  $-$ & $-$ \\[3pt]
  $R_d/R_{\text{star}}$ & 
  $\mathcal{U}[0.00,\,0.05]$ & $0.03048 \pm 0.00043$ & 
  $-$ & $-$ \\[3pt]
  $e_d$ & 
  $\mathcal{F}[0]$ & $0$ & 
  $\mathcal{F}[0]$ & $0$ \\[3pt]
  $\omega_d$ [deg] & 
  $\mathcal{F}[90]$ & $90$ & 
  $\mathcal{F}[90]$ & $90$ \\[3pt]
  $K_d$ [m s$^{-1}$] & 
  $-$ & $-$ & 
  $\mathcal{U}[0,\,10]$ & $1.45^{+0.52}_{-0.48}$ \\[3pt]
  \multicolumn{5}{c}{} \\[3pt]
  
  \textbf{GP's hyperparameters} & \multicolumn{4}{c}{} \\[3pt]
  $P_{\text{rot}}$ [days] & 
  $\mathcal{U}[8,\,12]$ & $9.613^{+0.055}_{-0.059}$ & 
  $\mathcal{U}[8,\,12]$ & $9.675 \pm 0.003$ \\[3pt]
  $\lambda_{\text{e, HARPS}}$ [days] & 
  $-$ & $-$ & 
  $\mathcal{U}[0,1000]$ & $21.1^{+2.0}_{-1.2}$ \\[3pt]
  $\lambda_{\text{e, HARPS-N}}$ [days] & 
  $-$ & $-$ & 
  $\mathcal{U}[0,1000]$ & $687^{+161}_{-117}$ \\[3pt]
  $\lambda_{\text{e, K2}}$ [days] & 
  $\mathcal{U}[0,1000]$ & $9.58^{+0.41}_{-0.45}$ & 
  $-$ & $-$ \\[3pt]
  $\lambda_{\text{e, TESS}}$ [days] & 
  $\mathcal{U}[0,1000]$ & $17.8^{+6.5}_{-4.7}$ & 
  $-$ & $-$ \\[3pt]
  $\lambda_{\text{P, HARPS}}$ & 
  $-$ & $-$ & 
  $\mathcal{U}[0,5]$ & $0.528^{+0.061}_{-0.059}$ \\[3pt]
  $\lambda_{\text{P, HARPS-N}}$ & 
  $-$ & $-$ & 
  $\mathcal{U}[0,5]$ & $0.984 \pm 0.062$ \\[3pt]
  $\lambda_{\text{P, K2}}$ & 
  $\mathcal{U}[0,5]$ & $0.321 \pm 0.012$ & 
  $-$ & $-$ \\[3pt]
  $\lambda_{\text{P, TESS}}$ & 
  $\mathcal{U}[0,5]$ & $0.570^{+0.120}_{-0.094}$ & 
  $-$ & $-$ \\[3pt]
  $h_{\text{K2}}$ & 
  $\mathcal{U}[0,\,5 \times 10^5]$ & $2990^{+288}_{-246}$ & 
  $-$ & $-$ \\[3pt]
  $h_{\text{TESS}}$ & 
  $\mathcal{U}[0,\,5 \times 10^5]$ & $2965^{+264}_{-231}$ & 
  $-$ & $-$ \\[3pt]
  $A_{\text{RV},\,{\text{HARPS}}}$ & 
  $-$ & $-$ & 
  $\mathcal{U}[-500,\,500]$ & $7.6^{+2.0}_{-1.8}$ \\[3pt]
  $B_{\text{RV},\,{\text{HARPS}}}$ & 
  $-$ & $-$ & 
  $\mathcal{U}[0,\,500]$ & $33.3^{+7.2}_{-5.9}$ \\[3pt]
  $A_{\text{FWHM},\,{\text{HARPS}}}$ & 
  $-$ & $-$ & 
  $\mathcal{U}[-500,\,500]$ & $0.0504^{+0.0110}_{-0.0097}$ \\[3pt]
  $B_{\text{FWHM},\,{\text{HARPS}}}$ & 
  $-$ & $-$ & 
  $\mathcal{F}[0]$ & $0$ \\[3pt]
  $A_{\text{H}\alpha,\,{\text{HARPS}}}$ & 
  $-$ & $-$ & 
  $\mathcal{U}[-500,\,500]$ & $0.0087^{+0.0020}_{-0.0016}$ \\[3pt]
  $B_{\text{H}\alpha,\,{\text{HARPS}}}$ & 
  $-$ & $-$ & 
  $\mathcal{F}[0]$ & $0$ \\[3pt]
  $A_{\text{RV},\,{\text{HARPS-N}}}$ & 
  $-$ & $-$ & 
  $\mathcal{U}[-500,\,500]$ & $87^{+23}_{-20}$ \\[3pt]
  $B_{\text{RV},\,{\text{HARPS-N}}}$ & 
  $-$ & $-$ & 
  $\mathcal{U}[-500,\,500]$ & $420 \pm 42$ \\[3pt]
  $A_{\text{BIS},\,{\text{HARPS-N}}}$ & 
  $-$ & $-$ & 
  $\mathcal{U}[-500,\,500]$ & $16^{+20}_{-15}$ \\[3pt]
  $B_{\text{BIS},\,{\text{HARPS-N}}}$ & 
  $-$ & $-$ & 
  $\mathcal{U}[-500,\,500]$ & $-392^{+47}_{-45}$ \\[3pt]
  $A_{\text{H}\alpha,\,{\text{HARPS-N}}}$ & 
  $-$ & $-$ & 
  $\mathcal{U}[-500,\,500]$ & $0.00337^{+0.00033}_{-0.00030}$ \\[3pt]
  $B_{\text{H}\alpha,\,{\text{HARPS-N}}}$ & 
  $-$ & $-$ & 
  $\mathcal{F}[0]$ & $0$ \\[3pt]
  \multicolumn{5}{c}{} \\[3pt]
  
  \textbf{Others parameters} & \multicolumn{4}{c}{} \\[3pt]    
  $\rho_{\text{star}}$ [$\rho_\odot$] &
  $\mathcal{N}[2.40,\,0.15]$ & $2.02 \pm 0.08$ & 
  $-$ & $-$ \\[3pt]
  $c_1$ & 
  $\mathcal{U}[0,\,1]$ & $0.72 \pm 0.13$ & 
  $-$ & $-$ \\[3pt]
  $c_2$ & 
  $\mathcal{U}[-1,\,1]$ & $0.17^{+0.18}_{-0.21}$ & 
  $-$ & $-$ \\[3pt]
  Offset RV HARPS & 
  $-$ & $-$ & 
  $\mathcal{U}[-10000,\,10000]$ & $0.0^{+2.6}_{-2.8}$ \\[3pt]
  Offset RV HARPS-N & 
  $-$ & $-$ & 
  $\mathcal{U}[-10000,\,10000]$ & $-103 \pm 47$ \\[3pt]
  Offset FWHM HARPS & 
  $-$ & $-$ & 
  $\mathcal{U}[-10000,\,10000]$ & $7.607 \pm 0.018$ \\[3pt]
  Offset BIS HARPS-N & 
  $-$ & $-$ & 
  $\mathcal{U}[-10000,\,10000]$ & $15^{+16}_{-33}$ \\[3pt]
  Offset H$\alpha$ HARPS & 
  $-$ & $-$ & 
  $\mathcal{U}[-10000,\,10000]$ & $0.0087^{+0.0020}_{-0.0016}$ \\[3pt]
  Offset H$\alpha$ HARPS-N & 
  $-$ & $-$ & 
  $\mathcal{U}[-10000,\,10000]$ & $-0.076^{+0.036}_{-0.029}$ \\[3pt]
  Normalisation factor K2 & 
  $\mathcal{U}[10^{-6}, 10^6]$ & $902034^{+738}_{-725}$ & 
  $-$ & $-$ \\[3pt]
  Normalisation factor TESS & 
  $\mathcal{U}[10^{-6}, 10^6]$ & $19371^{+80}_{-111}$ & 
  $-$ & $-$ \\[3pt]
  Jitter RV HARPS & 
  $-$ & $-$ & 
  $\mathcal{U}[0.01,\,400.00]$ & $2.81^{+0.40}_{-0.37}$ \\[3pt]
  Jitter RV HARPS-N & 
  $-$ & $-$ & 
  $\mathcal{U}[0.01,\,400.00]$ & $7.76^{+1.00}_{-0.86}$ \\[3pt]
  Jitter FWHM HARPS & 
  $-$ & $-$ & 
  $\mathcal{U}[0.01,\,400.00]$ & $0.01550^{+0.00110}_{-0.00090}$ \\[3pt]
  Jitter BIS HARPS-N & 
  $-$ & $-$ & 
  $\mathcal{U}[0.01,\,400.00]$ & $10.3 \pm 1.4$ \\[3pt]
  Jitter H$\alpha$ HARPS & 
  $-$ & $-$ & 
  $\mathcal{U}[0.0001,\,100.0000]$ & $0.00235 \pm 0.00015$ \\[3pt]
  Jitter H$\alpha$ HARPS-N & 
  $-$ & $-$ & 
  $\mathcal{U}[0.0001,\,100.0000]$ & $0.00337^{+0.00033}_{-0.00030}$ \\[3pt]
  Jitter K2 & 
  $\mathcal{U}[0.01, 2000]$ & $62.7 \pm 1.6$ & 
  $-$ & $-$ \\[3pt]
  Jitter TESS & 
  $\mathcal{U}[0.01, 2000]$ & $3.89^{+0.39}_{-0.37}$ & 
  $-$ & $-$ \\[3pt]

\end{longtable}

\nolinenumbers
\tablefoot{Uncertainties correspond to the 16th and 84th percentiles of the posterior distributions. Reported values refer to the circular-orbit configuration.}

\endgroup

\twocolumn

\end{appendix}

\end{document}